\documentclass[11pt]{article}

\usepackage{smoothing_paper}

\usepackage[section]{placeins}
\usepackage{tabularx,ragged2e,booktabs,caption}

\newcommand{\expt}[1]{\mathrm{E}\left[#1\right]}

\newcommand{\rset}{\mathbb{R}}
\newcommand{\nset}{\mathbb{N}}
\newcommand{\zset}{\mathbb{Z}}

\newcommand{\PERIOD}{.}
\newcommand{\COMMA}{,}

\newcommand{\Ordo}[1]{{\mathcal{O}}\left(#1\right)}

\makeatletter
\def\BState{\State\hskip-\ALG@thistlm}
\makeatother

\title{Hierarchical adaptive sparse  grids and quasi Monte Carlo for option pricing under the rough Bergomi model}

\author{Christian Bayer\thanks{
 Weierstrass Institute for Applied Analysis and Stochastics (WIAS),
 Berlin, Germany.}
        \and Chiheb Ben Hammouda\thanks{King Abdullah University of Science and Technology (KAUST), Computer, Electrical and Mathematical Sciences \& Engineering Division (CEMSE), Thuwal $23955-6900$, Saudi Arabia ({\tt chiheb.benhammouda@kaust.edu.sa}).} 
\and  Ra\'ul Tempone\thanks{King Abdullah University of Science and Technology (KAUST), Computer, Electrical and Mathematical Sciences \& Engineering Division (CEMSE), Thuwal $23955-6900$, Saudi Arabia ({\tt raul.tempone@kaust.edu.sa}).} \thanks{Alexander von Humboldt Professor in Mathematics for Uncertainty Quantification, RWTH Aachen University, Germany.}}

\begin{document}
	\date{}
\maketitle
\begin{abstract}
The rough Bergomi (rBergomi) model, introduced recently in  \cite{bayer2016pricing}, is a promising rough volatility model in quantitative finance. It is a parsimonious model depending on only three parameters, and yet remarkably fits with empirical implied volatility surfaces. In the absence of analytical European option pricing methods for the model, and due to the non-Markovian nature of the fractional driver, the prevalent option is to use the Monte Carlo (MC) simulation for pricing. Despite recent advances in the MC method in this context, pricing under the rBergomi model is still a time-consuming task. To overcome this issue, we have designed a novel,  hierarchical approach, based on i) adaptive sparse grids quadrature (ASGQ), and ii) quasi Monte Carlo (QMC). Both techniques are coupled with a Brownian bridge construction and a Richardson extrapolation on the weak error. By uncovering the available regularity,  our hierarchical methods demonstrate substantial computational gains with respect to the standard MC method, when reaching a sufficiently small relative error tolerance in the price estimates across different parameter constellations, even for very small values of the Hurst  parameter. Our work opens a new research direction in this field, i.e., to investigate the performance of  methods  other than Monte Carlo for pricing and calibrating under the rBergomi model.

\

\end{abstract}

\thispagestyle{plain}

\setcounter{tocdepth}{1}

 \section{Introduction}
Modeling volatility to be stochastic, rather than deterministic as in the Black-Scholes model, enables quantitative analysts to  explain certain phenomena observed in option price data, in particular the implied volatility smile. However, this family of models has a  main drawback in failing  to capture the true steepness of the implied volatility smile close to maturity. Jumps can be added to stock price models to overcome this undesired feature, for instance by modeling the stock price process as an exponential L\'evy process. However, the addition of jumps to stock price processes remains controversial \cite{christensen2014fact,bajgrowicz2015jumps}. 

Motivated by the statistical analysis of realized volatility by Gatheral, Jaisson and Rosenbaum \cite{gatheral2018volatility} and the theoretical results on implied volatility    \cite{alos2007short,fukasawa2011asymptotic}, rough stochastic volatility has emerged as a new paradigm in quantitative finance, overcoming the observed limitations of  diffusive stochastic volatility models. In these models, the trajectories of the volatility  have lower H\"older regularity than the trajectories of standard Brownian motion \cite{bayer2016pricing,gatheral2018volatility}. In fact, they are based on fractional Brownian motion (fBm), which  is a centered Gaussian process whose covariance structure depends on  the so-called Hurst parameter, $H$ (we refer to  \cite{mandelbrot1968fractional,coutin07introduction,biagini2008stochastic} for more details regarding the fBm processes). In the rough volatility case, where $0<H<1/2$, the fBm has negatively correlated increments and rough sample paths.   Gatheral, Jaisson, and Rosenbaum \cite{gatheral2018volatility}  empirically demonstrated the advantages of such models. For instance, they showed that the log-volatility in practice has a similar behavior to  fBm with the Hurst exponent $H \approx 0.1$ at any reasonable time scale (see also  \cite{gatheral2014volatility_2}).  These results were confirmed  by Bennedsen, Lunde and Pakkanen \cite{bennedsen2016decoupling}, who studied over a thousand individual US equities and showed that $H$ lies in $(0,1/2)$ for each equity. Other  works \cite{bennedsen2016decoupling,bayer2016pricing,gatheral2018volatility} showed additional  benefits of  such rough volatility models over  standard stochastic volatility models,   in terms of explaining crucial phenomena  observed in  financial markets.

The rough Bergomi (rBergomi) model, proposed by Bayer, Friz and Gatheral \cite{bayer2016pricing}, was one of the first developed rough volatility models. This model, which depends on only three parameters, shows a remarkable fit to empirical implied volatility surfaces. The construction of the rBergomi model was performed by  moving from a physical to a pricing measure and by simulating prices under that model to fit  the implied volatility surface well in the case of the S\&P $500$ index with few parameters. The model may be seen as a non-Markovian extension of the Bergomi variance curve model \cite{bergomi2005smile}.

Despite the promising features of the rBergomi model, pricing  and hedging under such a model still constitutes a challenging and time-consuming task due  to the non-Markovian nature of the fractional driver.  In fact,  standard numerical pricing methods, such as PDE discretization schemes, asymptotic expansions and transform
methods, although efficient in the case of diffusion, are not easily  carried over to the rough setting (with the remarkable exception of the rough Heston model \cite{el2018roughening,el2018perfect,abi2019lifting,el2019characteristic} and its affine extensions \cite{jaber2019affine,gatheral2019affine}). Furthermore,  due to the lack of Markovianity and affine structure, conventional analytical pricing methods  do not apply. To the best of our knowledge, the only prevalent method for pricing  options under such models is the Monte Carlo (MC) simulation. In particular,  recent advances in simulation methods for the rBergomi model and different variants of pricing methods based on  MC under such a model   have been proposed in \cite{bayer2016pricing,bayer2017regularity,bennedsen2017hybrid,mccrickerd2018turbocharging,jacquier2018vix}.  For instance, in \cite{mccrickerd2018turbocharging}, the authors use a novel composition of variance reduction methods. When pricing under the rBergomi model, they achieved substantial computational gains  over the standard MC method.  Greater  analytical understanding of option pricing and implied volatility under this model has been achieved  in \cite{jacquier2017pathwise,forde2017asymptotics,bayer2018short}.  It is crucial to note that hierarchical variance reduction methods, such
as multilevel Monte Carlo (MLMC), are inefficient in this context, because of the poor behavior of the strong error, that is of the order of $H$ \cite{neuenkirch2016order}.

Despite recent advances in the MC method, pricing under the rBergomi model is still computationally expensive. To overcome this issue,  we design  novel, fast option pricers  for options whose underlyings follow the rBergomi model,  based on i) adaptive sparse grids quadrature (ASGQ), and ii) quasi Monte Carlo (QMC). Both techniques are coupled with Brownian bridge construction and Richardson extrapolation.  To use these two deterministic quadrature techniques (ASGQ and QMC) for our purposes, we  solve two main issues that constitute the two stages of our newly designed method. In the first stage, we smoothen the integrand by using the conditional expectation tools, as was proposed in \cite{romano1997contingent} in the context of Markovian stochastic volatility  models, and in \cite{bayersmoothing} in the context of basket options.   In a second stage, we apply the deterministic quadrature method, to solve the integration problem. At this stage, we apply two hierarchical representations,  before using the ASGQ or QMC method, to overcome the issue of facing a high-dimensional integrand due to the discretization scheme used for simulating the rBergomi dynamics. Given that ASGQ and QMC benefit from anisotropy, the first representation consists in applying a hierarchical  path generation method, based on a Brownian
bridge construction, with the aim of reducing the effective dimension. The second technique consists in applying Richardson extrapolation to reduce the bias (weak error), which, in turn, reduces the  number of time steps needed in the coarsest level to achieve a certain error tolerance and consequently  the maximum number of dimensions needed for the integration problem. We emphasize that we are interested in  the pre-asymptotic regime (corresponding to a small number of time steps), and that the use of Richardson extrapolation is justified by  conjectures  \ref{conj: Weak error structure} and \ref{conj: Richardson extrapol}, and  our observed experimental results in that regime,  which suggest, in particular, that we have convergence of order one for the weak error and  that the pre-asymptotic regime is enough to achieve sufficiently accurate estimates for the option prices. Furthermore, we stress that no proper weak error analysis has been performed in the rough volatility context, which proves to be subtle, due to the  absence of a Markovian structure. We also believe that we are the first to claim that both hybrid and exact schemes have  a weak error  of order one, which is justified at least numerically.

To the best of our knowledge, we are also the first to propose (and design) a pricing method, in the context of rough volatility models, one that is based on deterministic quadrature methods. As illustrated by our numerical experiments for  different parameter constellations, our proposed methods appear to be competitive to the MC approach, which is the only prevalent method in this context.  Assuming one targets price estimates with a sufficiently small relative error tolerance, our proposed methods demonstrate substantial computational gains  over the standard MC method, even for very small values of  $H$. However, we do not claim that these gains will hold in the asymptotic regime, which requires a  higher accuracy. Furthermore,  in this work, we limit ourselves to comparing our novel proposed methods against the standard MC. A more systematic comparison with the variant of MC proposed in \cite{mccrickerd2018turbocharging} is left for future research. 

We emphasize that  applying deterministic  quadrature for the family of (rough) stochastic  volatility models is not straightforward, but we propose an original way  to overcome the high dimensionality of the integration domain, by first reducing the total dimension using Richardson extrapolation and then coupling Brownian bridge construction with ASGQ or QMC for an optimal performance of our proposed methods. Furthermore, our proposed methodology shows a robust performance with respect to the values of the Hurst parameter $H$, as  illustrated through the different numerical examples with even very low values of $H$. We stress that the performance of  hierarchical variance reduction methods such as MLMC  is very sensitive to the values of $H$, since  rough volatility models are numerically tough in the sense that they admit low convergence rates for the mean square approximation (strong convergence rates) (see \cite{neuenkirch2016order}).

While our focus is on the rBergomi model, our approach is applicable to a wide class of stochastic volatility models and in particular rough volatility models, since the approach that we propose does not use any particular property of the rBergomi model, and even the analytic smoothing step can be performed in almost similar manner for any stochastic volatility model.

This paper is structured as follows: We begin in Section \ref{sec:Problem setting} by  introducing  the pricing framework that we are considering in this study. We provide some details about the rBergomi model, option pricing under this model and the simulation schemes used to simulate asset prices following the rBergomi dynamics. We also  explain how we choose the optimal simulation scheme for an optimal performance of our approach. In Section \ref{sec:Weak error analysis}, we discuss the weak error in the context of the rBergomi. Then, in Section \ref{sec:Details our approach and error bounds}, we explain the different building blocks that constitute our proposed methods, which are basically ASGQ, QMC, Brownian bridge construction, and Richardson extrapolation. Finally, in Section \ref{sec:Numerical tests} we show the results obtained through the different numerical experiments conducted across different parameter constellations for the rBergomi model. The reported results show the promising potential of our proposed methods in this context.

 \section{Problem setting}\label{sec:Problem setting}
In this section, we introduce the pricing framework that we consider in this work. We start  by giving some details on the rBergomi model proposed in \cite{bayer2016pricing}. We then derive the formula of the price of a European call option under the rBergomi model in Section \ref{sec:Option pricing under rBergomi model}. Finally, we explain  some details about the schemes that we use to simulate the dynamics of asset prices under the rBergomi model.

\subsection{The rBergomi model}\label{sec:The rBergomi model}

We consider the rBergomi model for the price process $S_t$ as defined in  \cite{bayer2016pricing}, normalized to $r=0$ ($r$ is the interest rate), which is defined by

\begin{align}\label{eq:rBergomi_model1}
	dS_t &= \sqrt{v_t} S_t dZ_t, \nonumber \\
	v_t &= \xi_0(t) \exp\left( \eta \widetilde{W}_t^H - \frac{1}{2} \eta^2 t^{2H} \right),
\end{align}
where the Hurst parameter $0 < H < 1/2$  and  $\eta>0$. We refer to $v_t$ as the variance process, and $\xi_0(t) = \expt{v_t}$ is  the forward variance curve.  Here, $\widetilde{W}^H $ is a certain Riemann-Liouville fBm
process \cite{marinucci1999alternative,picard2011representation},  defined by
\begin{align}\label{eq:Volterra process}
	\widetilde{W}_t^H = \int_0^t K^H(t-s) dW_s^1, \quad t \ge 0 \COMMA
\end{align}
where the kernel $K^H : \rset_+  \rightarrow \rset_+$ is
\begin{equation*}
 \quad K^H(t-s) = \sqrt{2H} (t-s)^{H - 1/2},\quad \forall \: 0 \le s \le t.
\end{equation*}
By construction, $\widetilde{W}^H $ is a centered, locally $(H-\epsilon)$- H\"older continuous Gaussian process with $\text{Var}\left[\widetilde{W}^H_t \right] = t^{2H}$, and a dependence structure defined by 
 \begin{equation*}
 \expt{\widetilde{W}^H_u  \widetilde{W}^H_v}=u^{2H} C\left(\frac{v}{u} \right),\quad v >u \COMMA
 \end{equation*}
 where for $x \ge 1$ and $\gamma=\frac{1}{2}-H$
 \begin{equation*}
C(x)=2H \int_{0}^1 \frac{ds}{(1-s)^{\gamma} (x-s)^{\gamma}}.
 \end{equation*}
In \eqref{eq:rBergomi_model1} and \eqref{eq:Volterra process}, $W^1, Z$ denote two \emph{correlated} standard Brownian motions with correlation $\rho \in ]-1,0]$, so that we can represent $Z$ in terms of $W^1$ as
\begin{align*}
	Z=\rho	W^1+ \bar{\rho}W^\perp = \rho W^1+\sqrt{1-\rho^2} W^\perp,
\end{align*}
where $(W^1,W^\perp)$ are two independent standard Brownian motions.
Therefore, the solution to \eqref{eq:rBergomi_model1}, with $S(0)=S_0$, can be written as 

\begin{align}\label{eq:rBergomi_model}
	S_t&= S_0  \operatorname{exp}\left( \int_{0}^{t} \sqrt{v(s)} dZ(s)- \frac{1}{2} \int_{0}^{t} v(s) ds   \right),\quad S_0>0 \nonumber\\
	v_u&=\xi_0(u) \operatorname{exp}\left( \eta \widetilde{W}_u^H- \frac{\eta^2}{2} u^{2H} \right), \quad \xi_0>0 \PERIOD
\end{align}
\subsection{Option pricing under the rBergomi model}\label{sec:Option pricing under rBergomi model}

We are interested in pricing European call options under the rBergomi model. Assuming $S_0 = 1$, and using the conditioning argument on the $\sigma$-algebra generated by $W^1$, and denoted by $\sigma(W^1(t) ,t \le T)$ (an argument first used by \cite{romano1997contingent} in the context of Markovian stochastic volatility  models), we can  show that the call price is given by
\begin{align}\label{BS_formula_rbergomi}
	C_{\text{RB}}\left( T, K \right) &= \text{E}\left[ \left(S_T - K \right)^+ \right]  \nonumber\\
	&=\expt{\expt{(S_T-K)^+ \mid \sigma(W^1(t) ,t \le T)}}\nonumber \\
	&=\text{E}\left[C_{\text{BS}}\left( S_0 = \operatorname{exp}\left(\rho \int_0^T \sqrt{v_t} dW_t^1 - \frac{1}{2}
	\rho^2 \int_0^T v_t dt\right),\ k = K, \ \sigma^2 = (1-\rho^2)
	\int_0^T v_t dt \right) \right],
\end{align}
where $C_{\text{BS}}(S_0,k,\sigma^2)$ denotes the Black-Scholes call price function, for initial spot price $S_0$, strike price $k$ and volatility $\sigma^2$.

\eqref{BS_formula_rbergomi} can be obtained by using the orthogonal decomposition of $S_t$ into $S_{t}^1$ and $S_{t}^2$, where
\begin{align*}
	S_t^1=\mathcal{E}\{ \rho \int_{0}^{t}  \sqrt{v_s} dW_s^1\}, \: S_t^2= \mathcal{E}\{ \sqrt{1-\rho^2} \int_{0}^{t}  \sqrt{v_s} dW_s^\perp  \}	,
\end{align*}
and $\mathcal{E}(.)$ denotes the stochastic exponential; then,  by conditional log-normality, we have
\begin{align*}
	\log S_t \mid \sigma\{ W^1(s), s\le t\} \sim \mathcal{N}\left( \log S_t^1-\frac{1}{2} (1-\rho^2) \int_{0}^{t} v_s ds , (1-\rho^2) \int_{0}^{t} v_s ds \right).
\end{align*}
We point out that the analytical smoothing, based on conditioning, performed in \eqref{BS_formula_rbergomi} enables us to uncover the available regularity, and hence  get a smooth, analytic integrand inside the expectation. Therefore, applying a deterministic quadrature technique such as ASGQ or QMC becomes an adequate option for computing the call price, as we will investigate later. A similar conditioning was used in \cite{mccrickerd2018turbocharging} but for variance reduction purposes only.

\subsection{Simulation of the rBergomi model}\label{sec:Simulation of the rBergomi model}

One of the numerical challenges encountered in the simulation of  rBergomi dynamics  is the computation of  $\int_{0}^{T} \sqrt{v_t} dW_t^1$ and $V=\int_{0}^{T} v_t dt$ in \eqref{BS_formula_rbergomi}, mainly because of the singularity of the Volterra kernel $K^H(s-t)$ at the diagonal $s = t$. In fact,  one needs to jointly simulate two Gaussian processes $(W_t^1, \widetilde{W}^H_t: 0 \le t \le T)$, resulting in $W^1_{t_1},\dots, W^1_{t_N}$ and $\widetilde{W}^H_{t_1},\dots, \widetilde{W}^H_{t_N}$ along a given time grid $t_1 <\dots < t_N$. In the literature, there are essentially two suggested ways to achieve this:
 \begin{enumerate}
 	\item[i)] \textbf{Covariance based approach (exact simulation) \cite{bayer2016pricing,bayer2018short}}: $W^1_{t_1},\dots, W^1_{t_N}, \widetilde{W}^H_{t_1},\dots, \widetilde{W}^H_{t_N}$ together form a ($2N$)-dimensional Gaussian random vector with a computable covariance matrix, and therefore one can use Cholesky decomposition of the covariance matrix to produce exact samples of $W^1_{t_1},\dots, W^1_{t_N}, \widetilde{W}^H_{t_1},\dots, \widetilde{W}^H_{t_N}$ from $2 N$-dimensional Gaussian random vector as  an input. This method is exact but slow. The simulation  requires $\Ordo{N^2}$ flops. Note that the offline cost is $\Ordo{N^3}$ flops.
 	
 	\item[ii)]  \textbf{The hybrid scheme of \cite{bennedsen2017hybrid}}: This scheme uses a different approach, which is essentially based on  Euler discretization, and approximates the kernel function in \eqref{eq:Volterra process} by a power function near zero and by a step function elsewhere, which results in an approximation  combining Wiener integrals of the power function and a Riemann sum (see \eqref{eq:Hybrid_scheme_pre} and \cite{bennedsen2017hybrid} for more details). This approximation is  inexact in the sense that samples produced here do not exactly have the distribution of $W^1_{t_1},\dots, W^1_{t_N}, \widetilde{W}^H_{t_1},\dots, \widetilde{W}^H_{t_N}$.  However, they are much more accurate than the samples produced from a simple Euler discretization, and much faster than method $(i)$. As in method $(i)$, in this case, we need a $2 N$-dimensional Gaussian random input vector to produce one 	sample of $W^1_{t_1},\dots, W^1_{t_N}, \widetilde{W}^H_{t_1},\dots, \widetilde{W}^H_{t_N}$.
 \end{enumerate} 

\subsubsection{On the choice of the simulation scheme in our approach}\label{sec: choice of simulation scheme}
\begin{figure}[h!]
	\centering
	\begin{subfigure}{.4\textwidth}
		\centering
		\includegraphics[width=1\linewidth]{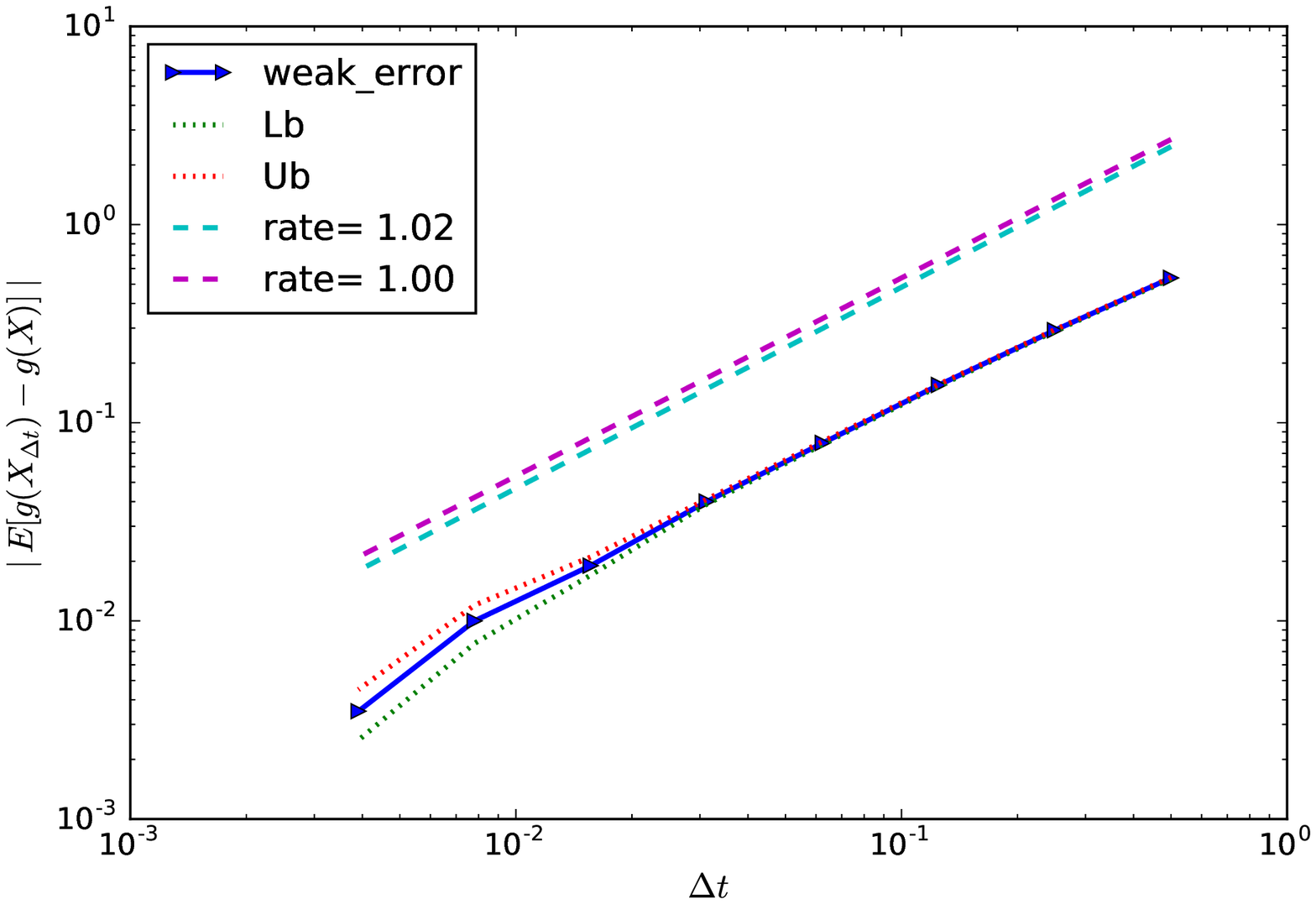}
		\caption{}
		\label{fig:set1_weak_rate_hybrid}
	\end{subfigure}%
	\begin{subfigure}{.4\textwidth}
		\centering
		\includegraphics[width=1\linewidth]{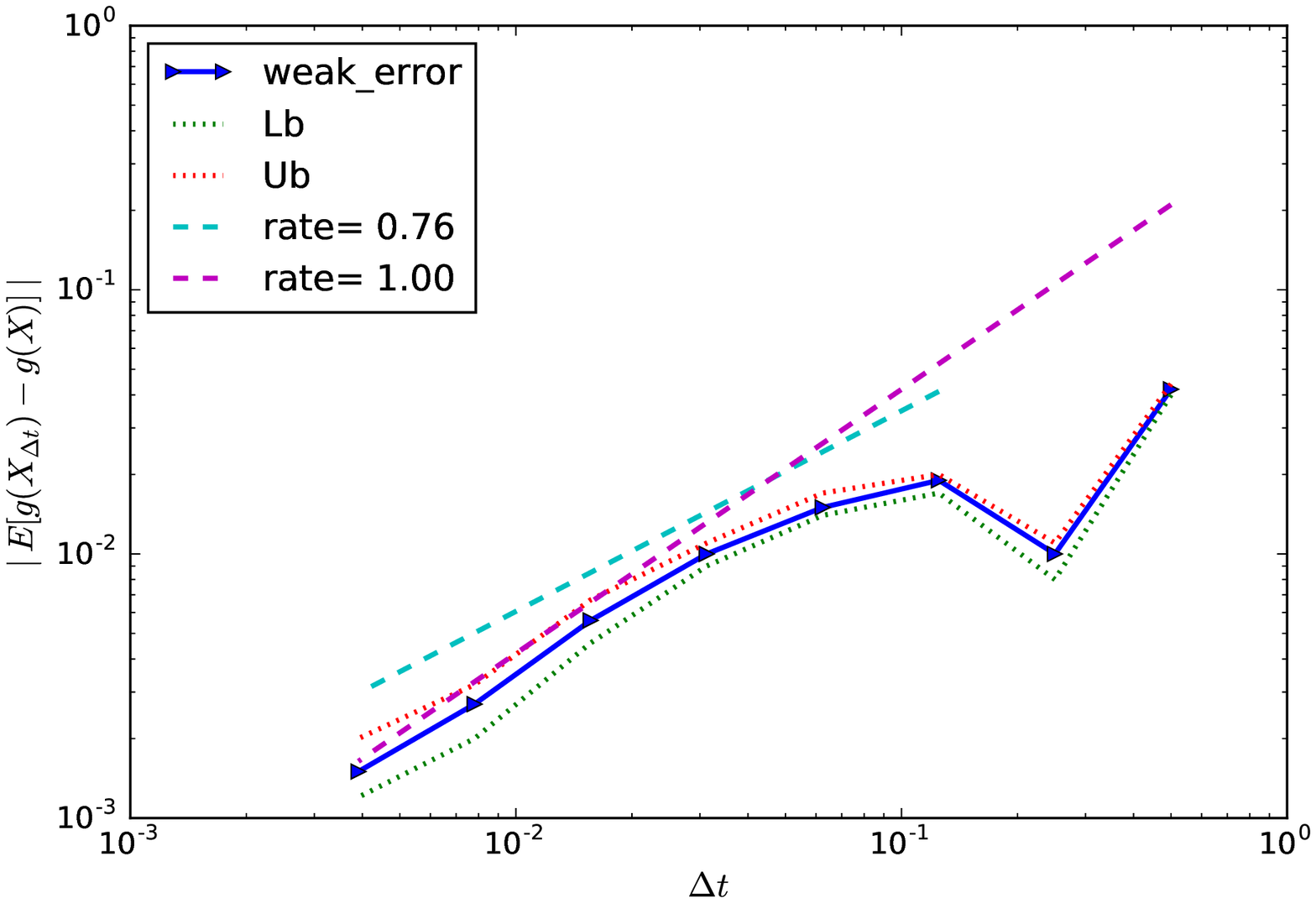}
		\caption{}
		\label{fig:set1_weak_rate_exact}
	\end{subfigure}
	\caption{The convergence of the weak error $\mathcal{E}_B$,  defined in \eqref{eq: Weak_error_hyb_chol}, using MC with $6 \times 10^6$ samples, for Set $1$ parameter in Table \ref{table:Reference solution, using MC with $500$ time steps, of Call option price under rBergomi model, for different parameter constellation.}. We refer to $C_{\text{RB}}$ (as in \eqref{BS_formula_rbergomi}) for $\expt{g(X)}$, and to $C_{\text{RB}}^{N}$ (as in \eqref{BS_formula_rbergomi_2}) for $\expt{g(X_{\Delta t})}$. The upper and lower bounds are $95\%$ confidence intervals. a) With the hybrid scheme  b) With the exact scheme.}
	\label{fig:Weak_rate_set1_set_2_without_rich_hyb+chol}
\end{figure}

The choice of the simulation scheme in our approach was based on the observed behavior of the weak rates. Through our numerical experiments (see Table \ref{table:Reference solution, using MC with $500$ time steps, of Call option price under rBergomi model, for different parameter constellation.} for the tested examples), we observe that, although the hybrid and exact schemes  seem to converge  with a  weak error of order $\Ordo{\Delta t}$, the pre-asymptotic behavior of the weak rate is different for both schemes (we provide a short discussion of the weak error in Section \ref{sec:Weak error analysis}). As an illustration, from Figure \ref{fig:Weak_rate_set1_set_2_without_rich_hyb+chol} for Set $1$ parameter in Table \ref{table:Reference solution, using MC with $500$ time steps, of Call option price under rBergomi model, for different parameter constellation.}, the hybrid scheme  has a consistent convergence behavior in the sense that it behaves in an asymptotic manner, basically right from the beginning, whereas the exact scheme does not. On the other hand, the constant  seems to be considerably smaller for the exact scheme. These two features make the hybrid scheme the  best choice   to work within our context, since our approach is based on hierarchical representations involving the use of Richardson extrapolation  (see Section \ref{sec:Richardson extrapolation}).
\FloatBarrier
\FloatBarrier

\subsubsection{The hybrid scheme}\label{sec: The hybrid scheme}
Let us denote  the number of time steps by $N$. As motivated in Section  \ref{sec: choice of simulation scheme}, in this work we use the hybrid scheme, which,  on an equidistant grid $\{0,\frac{1}{N},\frac{2}{N},\dots,\frac{NT}{N}\}$, is given by 
the following,
\begin{align}\label{eq:Hybrid_scheme_pre}
\widetilde{W}^H_{\frac{i}{N}} \approx \bar{W}^H_{\frac{i}{N}}&= \sqrt{2H} \left(  \sum_{k=1}^{\min(i,\kappa)} \int_{\frac{i}{N}-\frac{k}{N}}^{\frac{i}{N}-\frac{k}{N}+\frac{1}{N}} \left(\frac{i}{N}-s\right)^{H - 1/2} dW^1_s+\sum_{k=\kappa+1}^{i} \left(\frac{b_k}{N}\right)^{H - 1/2}  \int_{\frac{i}{N}-\frac{k}{N}}^{\frac{i}{N}-\frac{k}{N}+\frac{1}{N}} dW^1_s \right) \COMMA
\end{align}
which results for $\kappa=1$  in \eqref{eq:Hybrid_scheme}
\begin{align}\label{eq:Hybrid_scheme}
\widetilde{W}^H_{\frac{i}{N}} \approx \bar{W}^H_{\frac{i}{N}}&= \sqrt{2H} \left(  W^2_i+\sum_{k=2}^{i} \left(\frac{b_k}{N}\right)^{H-\frac{1}{2}} \left(W_{\frac{i-(k-1)}{N}}^1-W_{\frac{i-k}{N}}^1\right)\right),\quad 1 \le i \le N \COMMA
\end{align}
where 
\begin{equation}\label{eq:Hybrid_scheme_details}
 W^2_i=\int_{\frac{i-1}{N}}^{\frac{i}{N}} (\frac{i}{N}-s)^{H-1/2}dW^1_s   ,\quad b_k=\left(\frac{k^{H+\frac{1}{2}}-(k-1)^{H+\frac{1}{2} }}{H+\frac{1}{2}}\right)^{\frac{1}{H-\frac{1}{2}}} \PERIOD
 \end{equation}
The sum in \eqref{eq:Hybrid_scheme} requires the most computational effort in the simulation. Given that \eqref{eq:Hybrid_scheme} can be seen as discrete convolution  (see \cite{bennedsen2017hybrid}), we employ the fast Fourier transform to evaluate it, which results in  $\Ordo{N \log N}$ floating point operations.

We note that the variates $\bar{W}_0^{H},\bar{W}_{\frac{1}{N}}^{H},\dots,\bar{W}_{\frac{[N T]}{N}}^{H}$ are  generated by sampling $N$ i.i.d draws from a $(\kappa+1)$-dimensional Gaussian distribution and computing a discrete convolution.  For the case of $\kappa=1$, we denote these pairs  of Gaussian random vectors from now on by $(\mathbf{W}^{(1)},\mathbf{W}^{(2)})$, and we refer to \cite{bennedsen2017hybrid} for more details.

 \section{Weak error discussion}\label{sec:Weak error analysis}
To the best of our knowledge, no proper weak error analysis has been done in the rough volatility context, which proves to be subtle in the  absence of a Markovian structure. However, we try in this Section  to shortly discuss it in the context of the rBergomi model.

In this work, we are interested in approximating $\expt{g(X_T)}$, where $g$ is some smooth function and $X$ is the asset price under the rBergomi dynamics such that $X_t=X_t(W^{(1)}_{[0,t]},\widetilde{W}_{[0,t]})$, where $W^{(1)}$ is standard Brownian motion and  $\widetilde{W}$ is the fractional Brownian motion as given by \eqref{eq:Volterra process}.  Then we can express the approximation of $\expt{g(X_T)}$  using the  hybrid and exact schemes as the following 
\begin{align*}
\expt{g\left(X_T\left(W^{(1)}_{[0,T]},\widetilde{W}_{[0,T]}\right)\right)} \approx \expt{g\left(\overline{X}_N\left(W^{(1)}_1,\dots,W^{(1)}_N, \overline{W}_1, \dots,\overline{W}_N\right)\right)} \quad \textbf{(Hybrid  scheme)},
\end{align*}
\begin{align*}
\expt{g\left(X_T\left(W^{(1)}_{[0,T]},\widetilde{W}_{[0,T]}\right)\right)} \approx \expt{g\left(\overline{X}_N\left(W^{(1)}_1,\dots,W^{(1)}_N, \widetilde{W}_1, \dots,\widetilde{W}_N\right)\right)} \quad \textbf{(Exact  scheme)},
\end{align*}
where $\overline{W}$ is the approximation of $\widetilde{W}$  as given by \eqref{eq:Hybrid_scheme_pre} and $\overline{X}_N$ is the approximation of $X$ using $N$ time steps. In the following, to simplify notation, let  $\overline{\mathbf{W}}=(\overline{W}_1,\dots,\overline{W}_N)$, $\mathbf{W}^{1}=(W^{(1)}_1,\dots,W^{(1)}_N)$ and $\widetilde{\mathbf{W}}=(\widetilde{W}_1,\dots,\widetilde{W}_N)$. Then, the use of Richardson extrapolation in our  methodology presented in Section \ref{sec:Details our approach and error bounds} is mainly justified  by the conjecture \ref{conj: Weak error structure}. 
\begin{conjecture}\label{conj: Weak error structure}
If we denote by $\mathcal{E}_{B}^{\text{Hyb}}$ and $\mathcal{E}_{B}^{\text{Chol}}$ the weak errors produced by the hybrid and Cholesky scheme respectively, then we have
\begin{align*}
\mathcal{E}_{B}^{\text{Chol}}=\Ordo{\Delta t}, \quad \mathcal{E}_{B}^{\text{Hyb}} =\Ordo{\Delta t}.
\end{align*}
\end{conjecture}
We motivate Conjecture \ref{conj: Weak error structure} by writing
\begin{small}
\begin{align}\label{eq: Weak_error_hyb_chol}
\mathcal{E}_{B}^{\text{Hyb}}&=\abs{\expt{g\left(X_T\left(W^{(1)}_{[0,T]},\widetilde{W}_{[0,T]}\right) \right)}-\expt{g\left(\overline{X}_N\left(\mathbf{W}^{1}, \overline{\mathbf{W}}\right) \right)}} \nonumber\\
&\le \abs{\expt{g\left(X_T\left(W^{(1)}_{[0,T]},\widetilde{W}_{[0,T]}\right)\right)}-\expt{g\left(\overline{X}_N\left(\mathbf{W}^{1}, \widetilde{\mathbf{W}}\right)\right)}}+ \abs{ \expt{g\left(\overline{X}_N\left(\mathbf{W}^{1}, \overline{\mathbf{W}}\right)\right)}- \expt{g\left(\overline{X}_N\left(\mathbf{W}^{1}, \widetilde{\mathbf{W}}\right)\right)}}\nonumber\\
&\le \mathcal{E}_{B}^{\text{Chol}}+ \abs{ \expt{g\left(\overline{X}_N\left(\mathbf{W}^{1}, \overline{\mathbf{W}}\right)\right)}- \expt{g\left(\overline{X}_N\left(\mathbf{W}^{1}, \widetilde{\mathbf{W}}\right)\right)}}.
\end{align}
\end{small}
From the construction of the Cholesky scheme, we expect that the weak error is purely the discretization error, that is
\begin{align*}
\mathcal{E}_{B}^{\text{Chol}}=\Ordo{\Delta t},
\end{align*}
as it was observed by our numerical experiments (for illustration see Figure \ref{fig:set1_weak_rate_exact} for the case of Set $1$  in Table \ref{table:Reference solution, using MC with $500$ time steps, of Call option price under rBergomi model, for different parameter constellation.}).  The second term in the right-hand side of  \eqref{eq: Weak_error_hyb_chol} is basically related to approximating the integral \eqref{eq:Volterra process}  by \eqref{eq:Hybrid_scheme}. From our numerical experiments, it seems that this term is  at least  of order  $\Delta t$  and that its rate of convergence is independent of $H$ (for illustration see Figure \ref{fig:set1_weak_rate_hybrid} for the case of Set $1$  in Table \ref{table:Reference solution, using MC with $500$ time steps, of Call option price under rBergomi model, for different parameter constellation.}).

 
\section{Details of our hierarchical methods}\label{sec:Details our approach and error bounds}
We recall that our goal is to compute the expectation in \eqref{BS_formula_rbergomi}, and we remind  from  Section \ref{sec: The hybrid scheme} that we need    $2N$-dimensional Gaussian inputs, $\left(\mathbf{W}^{(1)},\mathbf{W}^{(2)}\right)$ for the used  hybrid  scheme ($N$ is the number of time steps in  the time grid). We can rewrite \eqref{BS_formula_rbergomi} as 
\begin{align}\label{BS_formula_rbergomi_2}
C_{\text{RB}}\left( T, K \right)&=\text{E}\left[C_{\text{BS}}\left( S_0 = \operatorname{exp}\left(\rho \int_0^T \sqrt{v_t} dW_t^1 - \frac{1}{2}
\rho^2 \int_0^T v_t dt\right),\ k = K, \ \sigma^2 = (1-\rho^2)
\int_0^T v_t dt \right) \right] \nonumber \\
&\approx \int_{\rset^{2N}} C_{BS} \left(G_{\text{rB}}(\mathbf{w}^{(1)},\mathbf{w}^{(2)})\right) \rho_{N}(\mathbf{w}^{(1)})  \rho_{N}(\mathbf{w}^{(2)}) d\mathbf{w}^{(1)} d\mathbf{w}^{(2)} \nonumber \\
&:=C_{\text{RB}}^{N},
\end{align}
where $G_{\text{rB}}$  maps  $2N$ independent standard Gaussian random inputs, formed by $\left(\mathbf{W}^{(1)},\mathbf{W}^{(2)}\right)$,  to the parameters fed to the Black-Scholes call price function, $C_{BS}$ (defined in \eqref{BS_formula_rbergomi}), and  $\rho_N$ is the multivariate Gaussian density, given by 
\begin{equation*}\label{eq: multivariate gaussian distribution}
\rho_N(\mathbf{z})=\frac{1}{(2 \pi)^{N/2}} e^{-\frac{1}{2} \mathbf{z}^T \mathbf{z}} \PERIOD
\end{equation*} 
Therefore, the initial integration problem that we are solving lives in $2N$-dimensional space, which becomes very large as the number of time steps $N$, used in the hybrid scheme, increases.

Our approach of approximating the expectation in \eqref{BS_formula_rbergomi_2} is based on hierarchical deterministic quadratures, namely  i) an ASGQ using the same construction as in \cite{haji2016multi} and ii) a randomized QMC based on lattice rules. In Section \ref{sec:Details of the MISC} we describe the ASGQ  method in our context, and in Section \ref{sec:Quasi Monte Carlo (QMC)} we provide details on the implemented QMC method.  To make an effective use of either the ASGQ or the QMC method, we  apply two techniques to overcome the issue of facing a high dimensional integrand due to the discretization scheme used for simulating the rBergomi dynamics. The first  consists in applying a hierarchical  path generation method, based on a Brownian
bridge construction, with the aim of reducing the effective dimension, as  described  in Section \ref{sec:Brwonian bridge construction}. The second technique consists in applying Richardson extrapolation to reduce the bias, resulting in reducing  the maximum number of dimensions needed for the integration problem. Details about  the Richardson extrapolation  are provided in Section \ref{sec:Richardson extrapolation}.

If we denote by $\mathcal{E}_{\text{tot}}$ the total error of approximating the  expectation in \eqref{BS_formula_rbergomi}, using the ASGQ estimator, $Q_N$ (defined by \eqref{eq:MISC_quad_estimator}), then we obtain a natural error decomposition
\begin{align}\label{eq:total_error_ASGQ}
\mathcal{E}_{\text{tot}} & \le \abs{C_{\text{RB}}-C_{\text{RB}}^N}+\abs{C_{\text{RB}}^N-Q_{N}} \le \mathcal{E}_B(N)+ \mathcal{E}_Q(\text{TOL}_{\text{ASGQ}},N),
\end{align}
where  $\mathcal{E}_Q$ is the quadrature error, $\mathcal{E}_B$  is the bias, $\text{TOL}_{\text{ASGQ}}$ is a user-selected tolerance for the ASGQ method, and $C_{\text{RB}}^N$ is the biased price computed with $N$ time steps, as given by \eqref{BS_formula_rbergomi_2}.

On the other hand, the total error of approximating the  expectation in \eqref{BS_formula_rbergomi} using the randomized QMC or MC estimator, $Q^{\text{MC(QMC)}}_N$ can be bounded by

\begin{align}\label{eq:total_error_MC}
	\mathcal{E}_{\text{tot}} & \le \abs{C_{\text{RB}}-C_{\text{RB}}^N}+\abs{C_{\text{RB}}^N-Q^{\text{MC (QMC)}}_N} \le \mathcal{E}_B(N)+ \mathcal{E}_{S}(M,N),
\end{align}
where  $\mathcal{E}_S$ is the statistical error\footnote{The statistical error estimate of MC or randomized QMC is  $C_{\alpha} \frac{\sigma_M}{\sqrt{M}}$, where $M$ is the number of samples and $C_{\alpha}=1.96$ for $95\%$ confidence interval.}, $M$ is the number of samples used for the MC or the randomized QMC method.
\begin{remark}
We note that, thanks to \eqref{BS_formula_rbergomi_2}, our approach, explained in the following sections, can be extended to any (rough)  stochastic volatility  dynamics, with the only difference of using another function instead  of   $G_{\text{rB}}$ in \eqref{BS_formula_rbergomi_2},  with specifics of the considered model embedded in this function.
\end{remark}
\subsection{Adaptive sparse grids quadrature (ASGQ)}\label{sec:Details of the MISC}

We assume that we want to approximate the expected value $\text{E}[f(Y)]$ of an analytic function $f\colon \Gamma \to \rset$ using a tensorization of quadrature formulas over $\Gamma$.

To introduce simplified notations, we start with the one-dimensional case. Let us denote by $\beta$ a non-negative integer, referred to as a ``stochastic discretization level", and by $m: \nset \rightarrow \nset$  a strictly increasing function with $m(0)=0$ and $m(1)=1$, that we call  ``level-to-nodes function". At level $\beta$, we consider a set of $m(\beta)$ distinct quadrature points in $\rset$, $\mathcal{H}^{m(\beta)}=\{y^1_\beta,y^2_\beta,\dots,y_\beta^{m(\beta)}\} \subset \rset$, and a set of quadrature weights, $\boldsymbol{\omega}^{m(\beta)}=\{\omega^1_\beta,\omega^2_\beta,\dots,\omega_\beta^{m(\beta)}\}$. We also let $C^0(\rset)$ be the set of real-valued continuous functions over $\rset$. We then define the quadrature operator as
\begin{equation*}
Q^{m(\beta)}:C^0(\rset) \rightarrow \rset, \quad Q^{m(\beta)}[f]= \sum_{j=1}^{m(\beta)} f(y^j_\beta) \omega_\beta^j.
\end{equation*}
In our case, we have in \eqref{BS_formula_rbergomi_2} a multi-variate integration problem with,  $f=C_{\text{BS}}\circ G_{\text{rB}}$, $\mathbf{Y}=(\mathbf{W}^{(1)},\mathbf{W}^{(2)})$, and  $\Gamma=\rset^{2N}$, in the previous notations. Furthermore, since we are dealing with Gaussian densities, using Gauss-Hermite quadrature points is the appropriate choice.

We define for any multi-index $\boldsymbol{\beta} \in \nset^{2N}$
$$Q^{m(\boldsymbol{\beta})}: C^0(\rset^{2N}) \rightarrow \rset,\quad  Q^{m(\boldsymbol{\beta})}= \bigotimes_{n = 1}^{2N} Q^{m(\beta_n)} \COMMA $$
where the $n$-th quadrature operator is understood to act only on the $n$-th variable of $f$. Practically, we obtain the value of $Q^{m(\boldsymbol{\beta})}[f]$  by using the grid $\mathcal{T}^{m(\boldsymbol{\beta})}= \prod_{n = 1}^{2N}  \mathcal{H}^{m(\beta_n)}$, with cardinality $\#\mathcal{T}^{m(\boldsymbol{\beta})}=\prod_{n=1}^{2N} m (\beta_n)$, and computing
$$ Q^{m(\boldsymbol{\beta})}[f]= \sum_{j=1}^{\#\mathcal{T}^{m(\boldsymbol{\beta})}} f(\hat{y}_j) \bar{\omega}_j \COMMA$$
where $\hat{y}_j \in \mathcal{T}^{m(\boldsymbol{\beta})}$ and $\bar{\omega}_j$ are  products of weights of the univariate quadrature rules. To simplify notation, hereafter, we replace  $Q^{m(\boldsymbol{\beta})}$ by $Q^{\boldsymbol{\beta}}$.

A direct approximation $\expt{f[\mathbf{Y}]} \approx Q^{\boldsymbol{\beta}}[f]$ is not an appropriate option,  due to the well-known ``curse of dimensionality". We use  a hierarchical ASGQ\footnote{More details about sparse grids can be found in \cite{bungartz2004sparse}.} strategy, specifically using the same
construction as in \cite{haji2016multi}, and which uses  stochastic discretizations  and a classic sparsification approach to obtain an effective approximation scheme for $\expt{f}$. 

\begin{figure}[htb]
 	\centering 
 	\begin{subfigure}{0.165\textwidth}
 		\includegraphics[width=\linewidth]{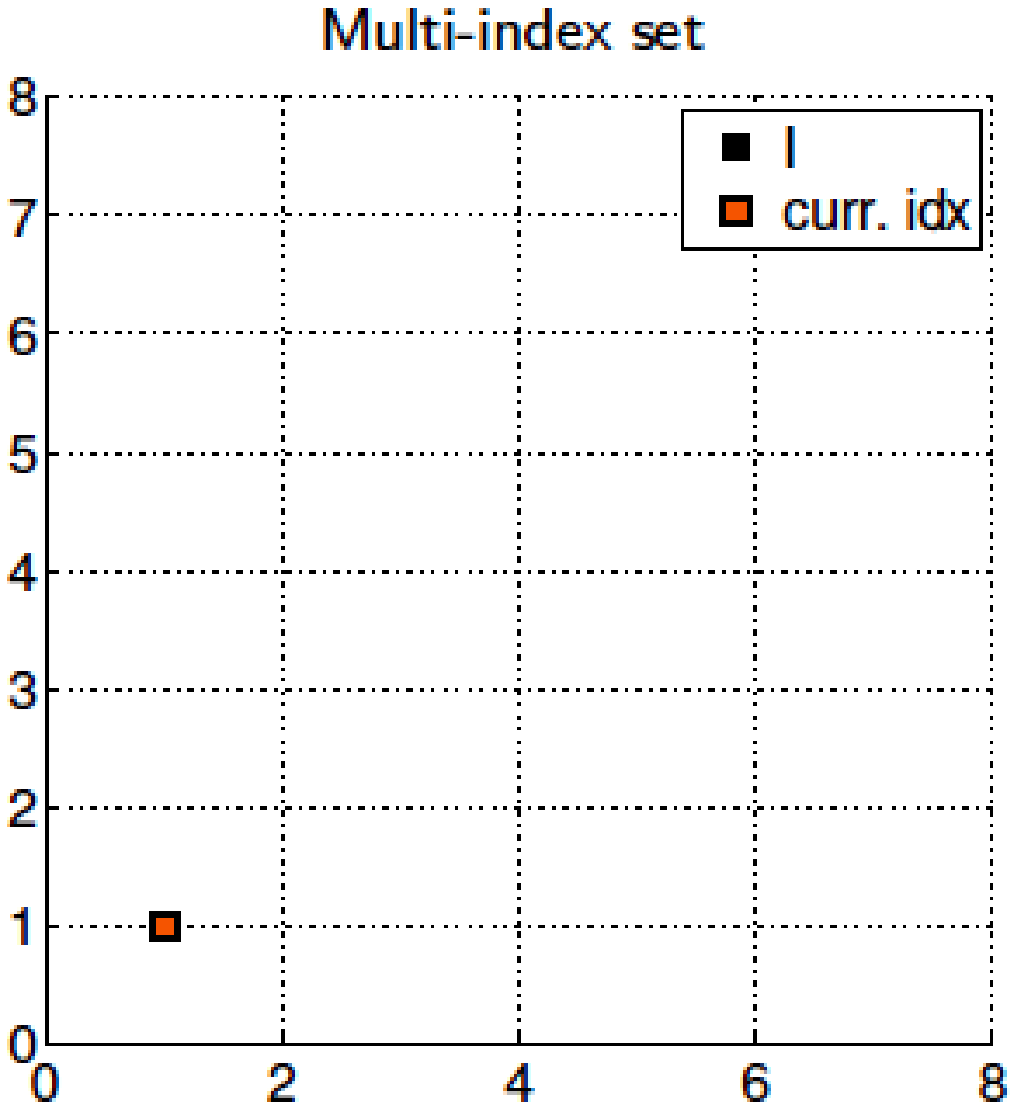}
 		\caption{}
 		\label{fig:1}
 	\end{subfigure}\hfil 
 	\begin{subfigure}{0.165\textwidth}
 		\includegraphics[width=\linewidth]{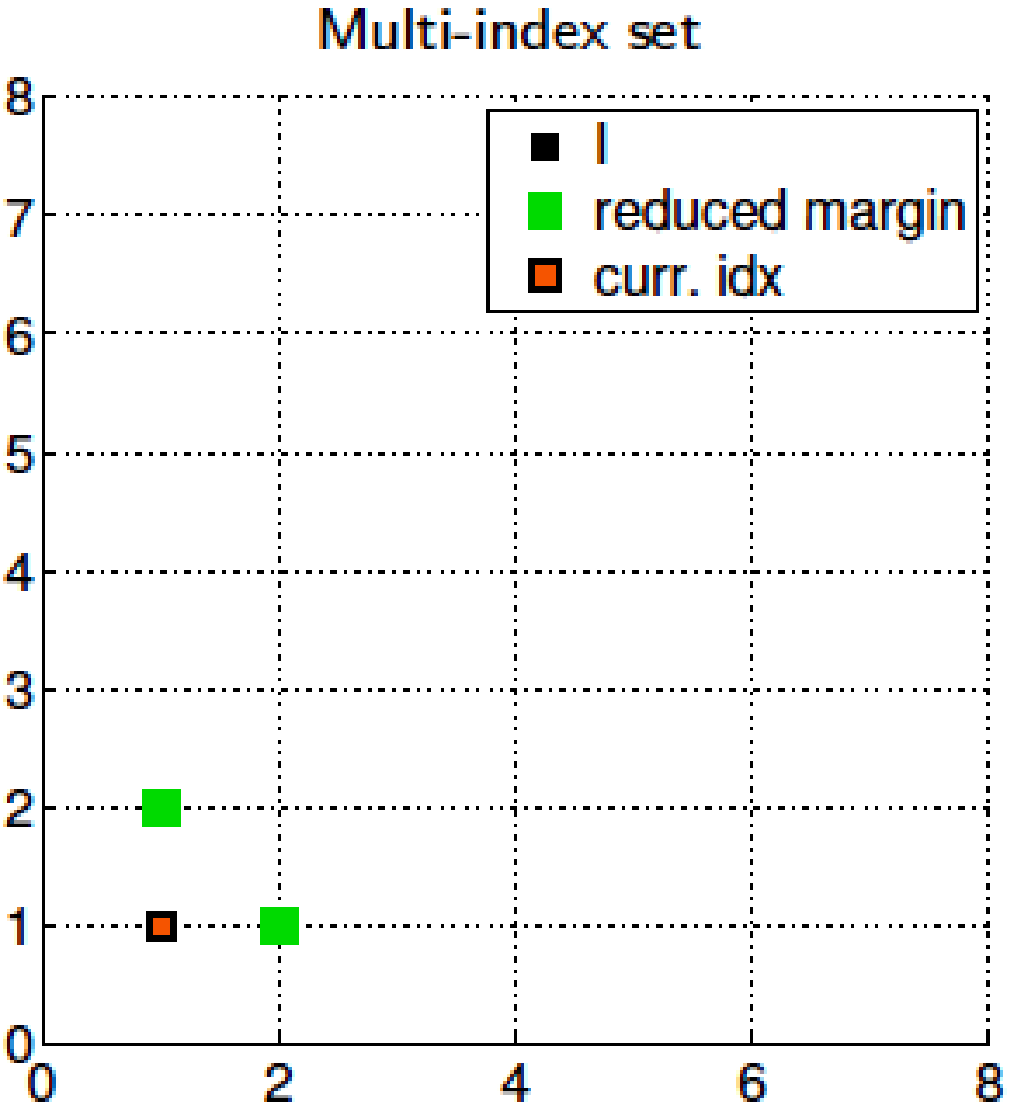}
 		\caption{}
 		\label{fig:2}
 	\end{subfigure}\hfil 
 	\begin{subfigure}{0.165\textwidth}
 		\includegraphics[width=\linewidth]{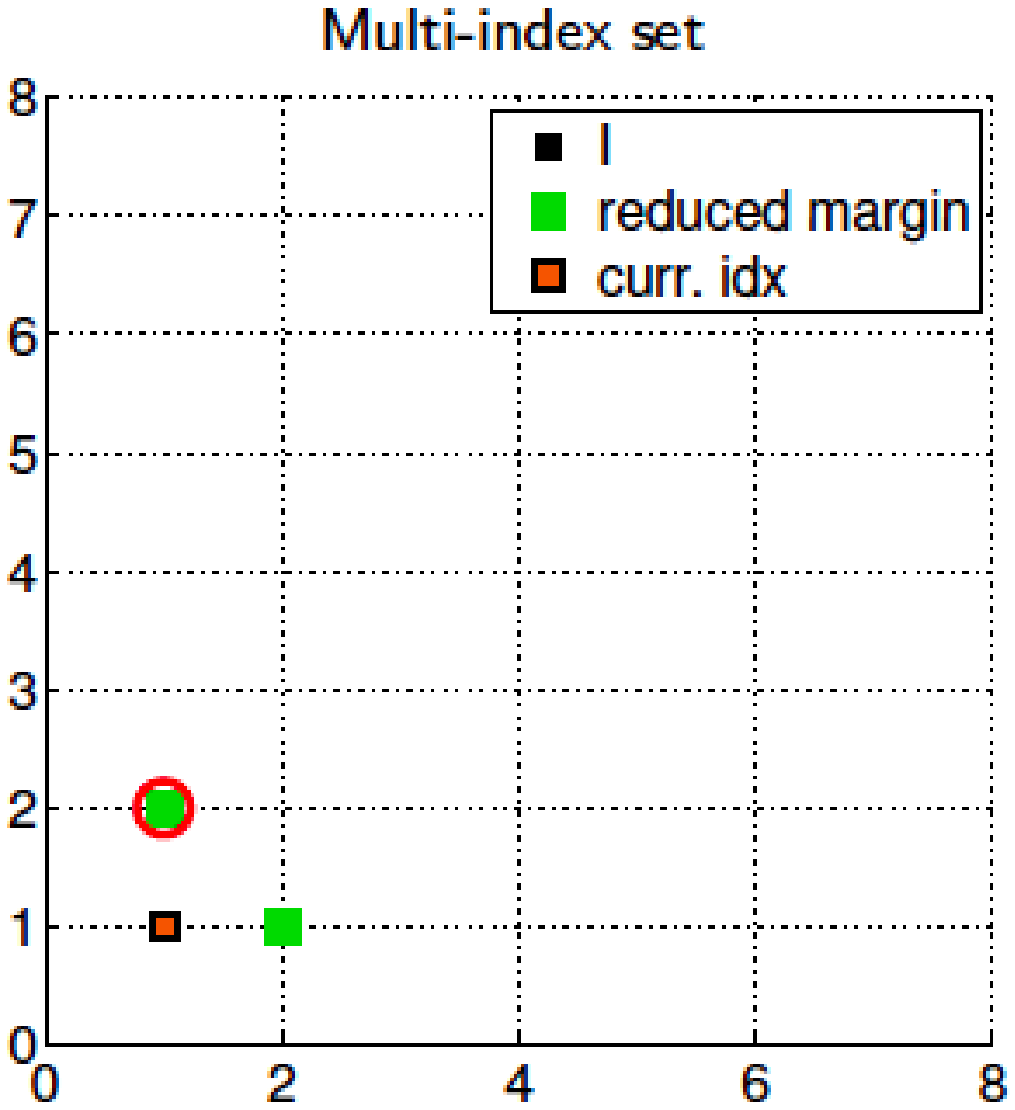}
 		\caption{}
 		\label{fig:3}
 	\end{subfigure}
 	\medskip
 	\begin{subfigure}{0.165\textwidth}
 		\includegraphics[width=\linewidth]{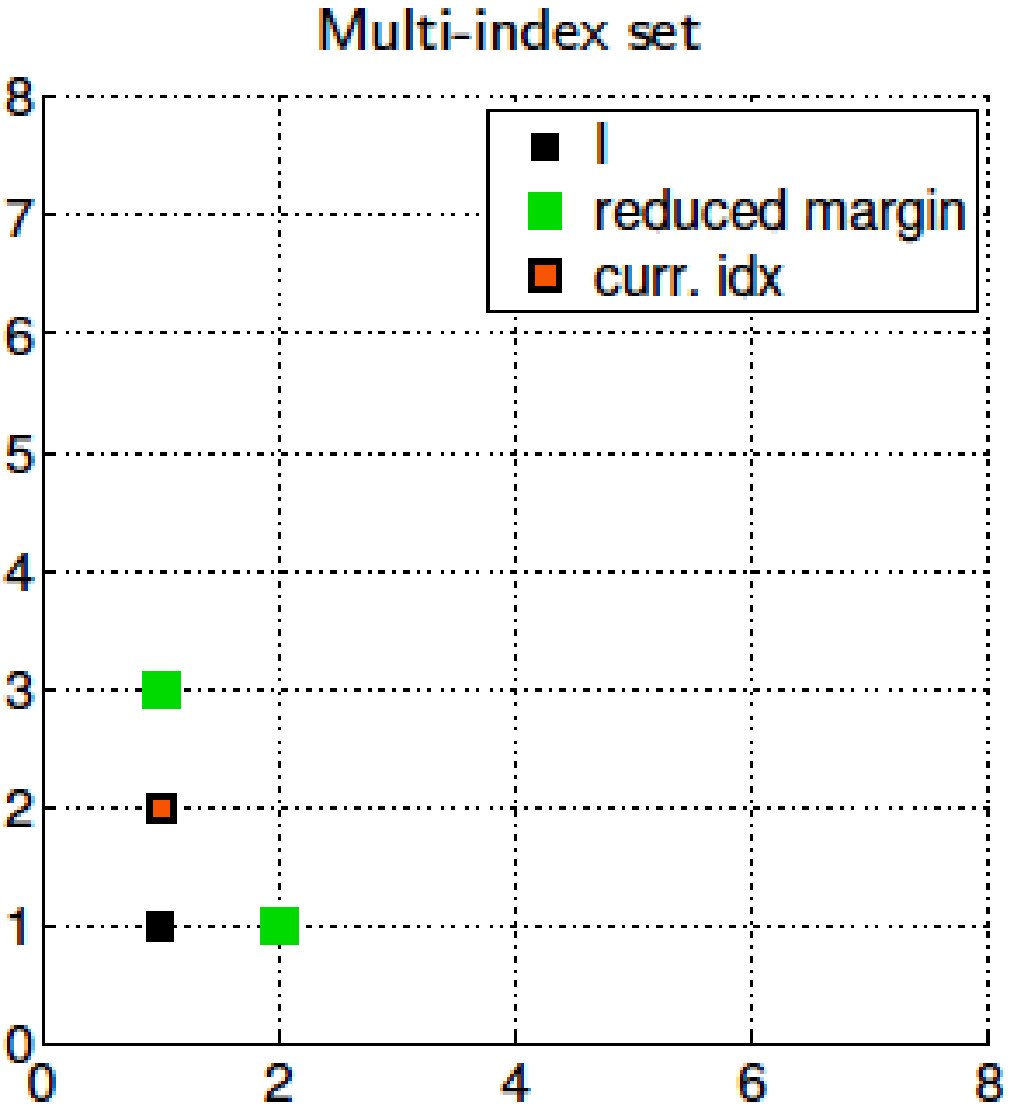}
 		\caption{}
 		\label{fig:4}
 	\end{subfigure}\hfil 
 	\begin{subfigure}{0.165\textwidth}
 		\includegraphics[width=\linewidth]{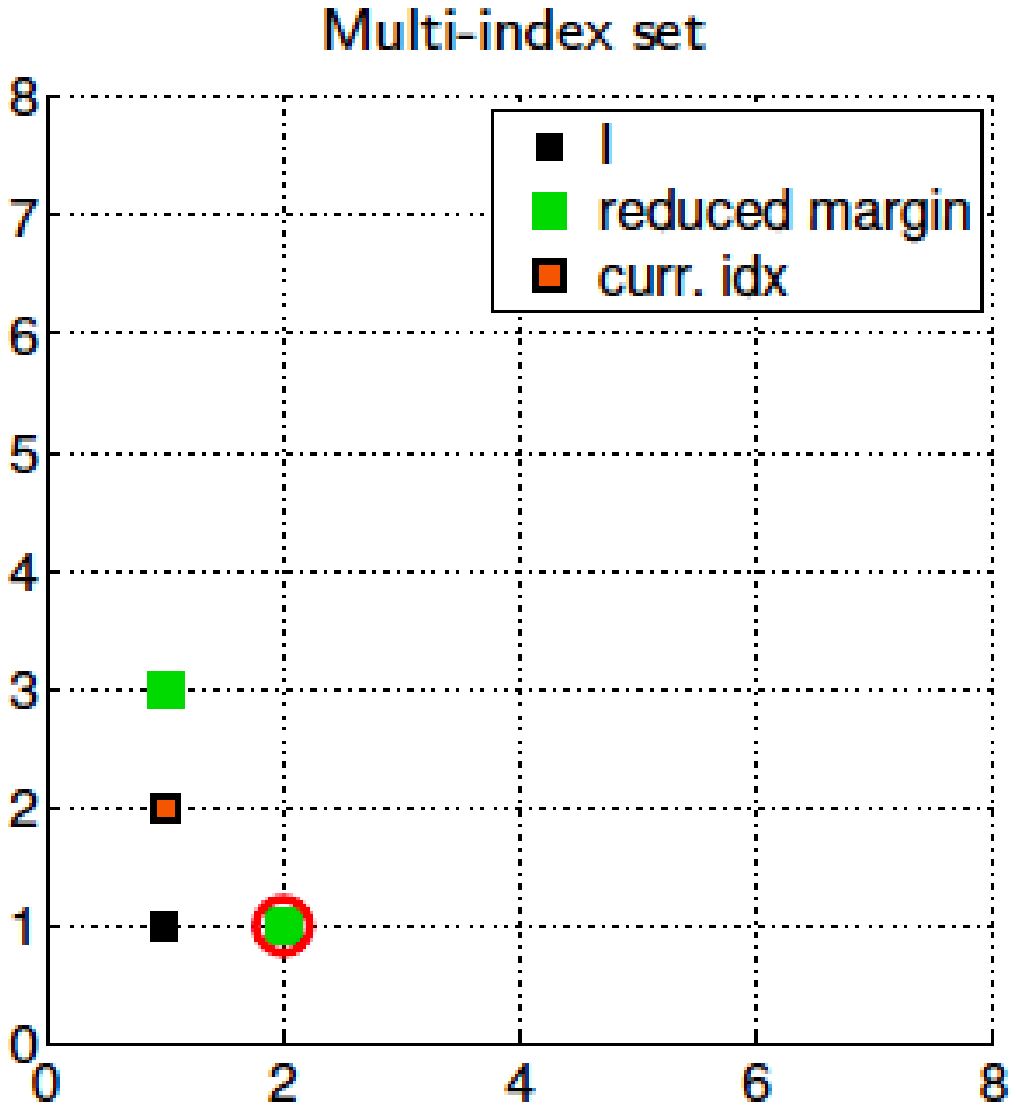}
 		\caption{}
 		\label{fig:5}
 	\end{subfigure}\hfil 
 	\begin{subfigure}{0.165\textwidth}
 		\includegraphics[width=\linewidth]{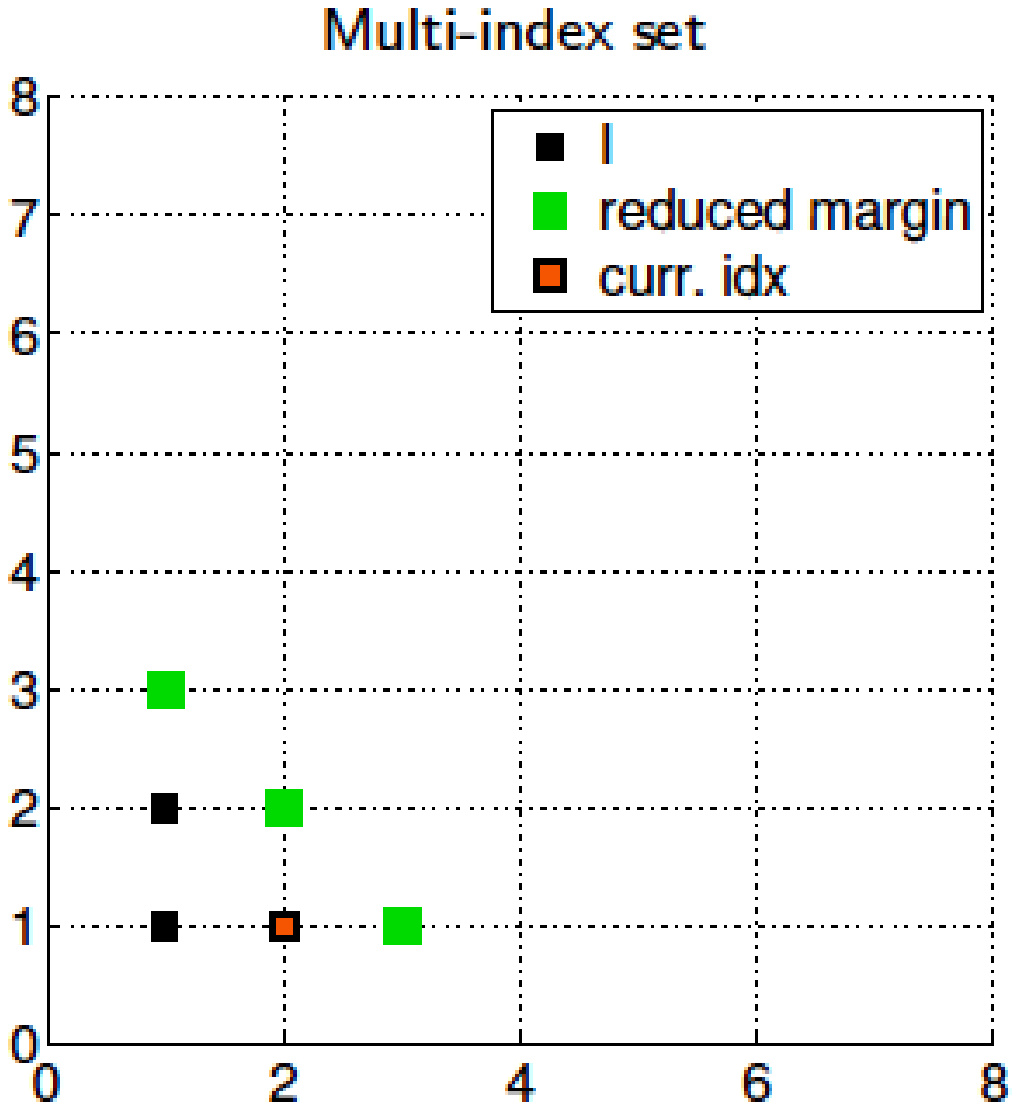}
 		\caption{}
 		\label{fig:6}
 	\end{subfigure}
 	\caption{Snapshots of the greedy construction of the index set for ASGQ method. A posteriori, adaptive construction: Given an index set $\mathcal{I}_k$, compute the profits of the neighbor indices using \eqref{eq:profit_equation}, and select the most profitable one.}
 	\label{fig:Construction of the index set for ASGQ method}
 \end{figure}

To be concrete, in our setting, we are left with a $2N$-dimensional Gaussian random input, which is chosen independently, resulting in  $2N$ numerical parameters for ASGQ, which we use as the basis of the multi-index construction. For a multi-index $\boldsymbol{\beta} = (\beta_n)_{n=1}^{2N} \in \mathbb{N}^{2N}$, we denote  by
$Q_N^{\boldsymbol{\beta}}$ the result of approximating \eqref{BS_formula_rbergomi_2} with a number of quadrature points  in the $i$-th dimension equal to  $m(\beta_i)$. We further define the set of
differences $\Delta Q_N^{\boldsymbol{\beta}}$ as follows: for a single index $1 \le i \le 2N$,
let
\begin{equation*}
\Delta_i Q_N^{\boldsymbol{\beta}} = \left\{ 
\aligned 
 Q_N^{\boldsymbol{\beta}} &- Q_N^{\boldsymbol{\beta}'}  \text{, with } \boldsymbol{\beta}' =\boldsymbol{\beta} - e_i, \text{ if } \boldsymbol{\beta}_i>0 \COMMA \\
 Q_N^{\boldsymbol{\beta}} &, \quad  \text{ otherwise,}
\endaligned
\right.
\end{equation*}
where $e_i$ denotes the $i$th $2N$-dimensional unit vector. Then, $\Delta
Q_N^{\boldsymbol{\beta}}$ is defined as
\begin{equation*}
\Delta Q_N^{\boldsymbol{\beta}} = \left( \prod_{i=1}^{2N} \Delta_i \right) Q_N^{\boldsymbol{\beta}}.
\end{equation*}
For instance, when $N = 1$, then 
\begin{multline*}
	\Delta Q_1^{\boldsymbol{\beta}} = \Delta_2 \Delta_1 Q_1^{(\beta_1, \beta_2)} = \Delta_2\left( Q_1^{(\beta_1,
		\beta_2)} - Q_1^{(\beta_1-1,\beta_2)} \right) = \Delta_2 Q_1^{(\beta_1,
		\beta_2)} - \Delta_2 Q_1^{(\beta_1-1,\beta_2)} 
	\\= Q_1^{(\beta_1, \beta_2)} - Q_1^{(\beta_1, \beta_2-1)} - Q_1^{(\beta_1-1, \beta_2)} + Q_1^{(\beta_1-1, \beta_2-1)}.
\end{multline*}
Given the definition of $C_{\text{RB}}^{N}$ by \eqref{BS_formula_rbergomi_2}, we have the telescoping property
\begin{equation*}
C_{\text{RB}}^{N}=Q_N^\infty = \sum_{\beta_1=0}^\infty \cdots \sum_{\beta_{2N} = 0}^\infty \Delta
Q_N^{(\beta_1, \ldots, \beta_{2N})} = \sum_{\boldsymbol{\beta} \in \mathbb{N}^{2N}} \Delta Q_N^{\boldsymbol{\beta}}.
\end{equation*}
The ASGQ estimator used for approximating \eqref{BS_formula_rbergomi_2}, and using a set of multi-indices $\mathcal{I}\subset \nset^{2N}$ is given by
\begin{equation}\label{eq:MISC_quad_estimator}
	Q_N^{\mathcal{I}} = \sum_{\boldsymbol{\beta} \in \mathcal{I}} \Delta Q_N^{\boldsymbol{\beta}}.
\end{equation}
The quadrature error in this  case  is given by
\begin{equation}\label{eq:quadrature error}
	\mathcal{E}_Q(\text{TOL}_{\text{ASGQ}},N) =\abs{Q_N^\infty - Q_N^\mathcal{I}} \le \sum_{\boldsymbol{\beta} \in \mathbb{N}^{2N} \setminus
		\mathcal{I}} \abs{\Delta Q_N^{\boldsymbol{\beta}}}.
\end{equation}
We define the work contribution, $\Delta \mathcal{W}_{\boldsymbol{\beta}}$, to be the computational cost  required to add  $\Delta Q_N^{\boldsymbol{\beta}}$ to $Q^{\mathcal{I}}_N$, and the error contribution, $\Delta E_{\boldsymbol{\beta}}$, to be  a measure of how much the quadrature error, defined in \eqref{eq:quadrature error}, would decrease once $\Delta Q_N^{\boldsymbol{\beta}}$  has been added to  $Q^{\mathcal{I}}_N$, that is 
\begin{align}\label{eq:Work_error_contributions}
\Delta E_{\boldsymbol{\beta}} &= \abs{Q^{\mathcal{I} \cup \{\boldsymbol{\beta}\}}_N-Q^{\mathcal{I}}_N}\\
\Delta \mathcal{W}_{\boldsymbol{\beta}} &= \text{Work}[Q^{\mathcal{I} \cup \{\boldsymbol{\beta}\}}_N]-\text{Work}[Q^{\mathcal{I}}_N].
\end{align}
 The  construction of the optimal  $\mathcal{I}$ is done by profit thresholding (see Figure \ref{fig:Construction of the index set for ASGQ method} for illustration), that is, for a certain threshold value $\bar{T}$, and a profit of a hierarchical surplus defined by
 \begin{equation}\label{eq:profit_equation}
 P_{\boldsymbol{\beta}}= \frac{\abs{\Delta E_{\boldsymbol{\beta}}}}{\Delta\mathcal{W}_{\boldsymbol{\beta}}},
 \end{equation}
the optimal index set  $\mathcal{I}$  for our ASGQ is given by 
 $\mathcal{I}=\{\boldsymbol{\beta} \in \nset^{2N}_+: P_{\boldsymbol{\beta}}	 \ge \bar{T}\}$. 
 \FloatBarrier
\FloatBarrier

\begin{figure}[h!]
	\centering
	\begin{subfigure}{.4\textwidth}
		\centering
		\includegraphics[width=1\linewidth]{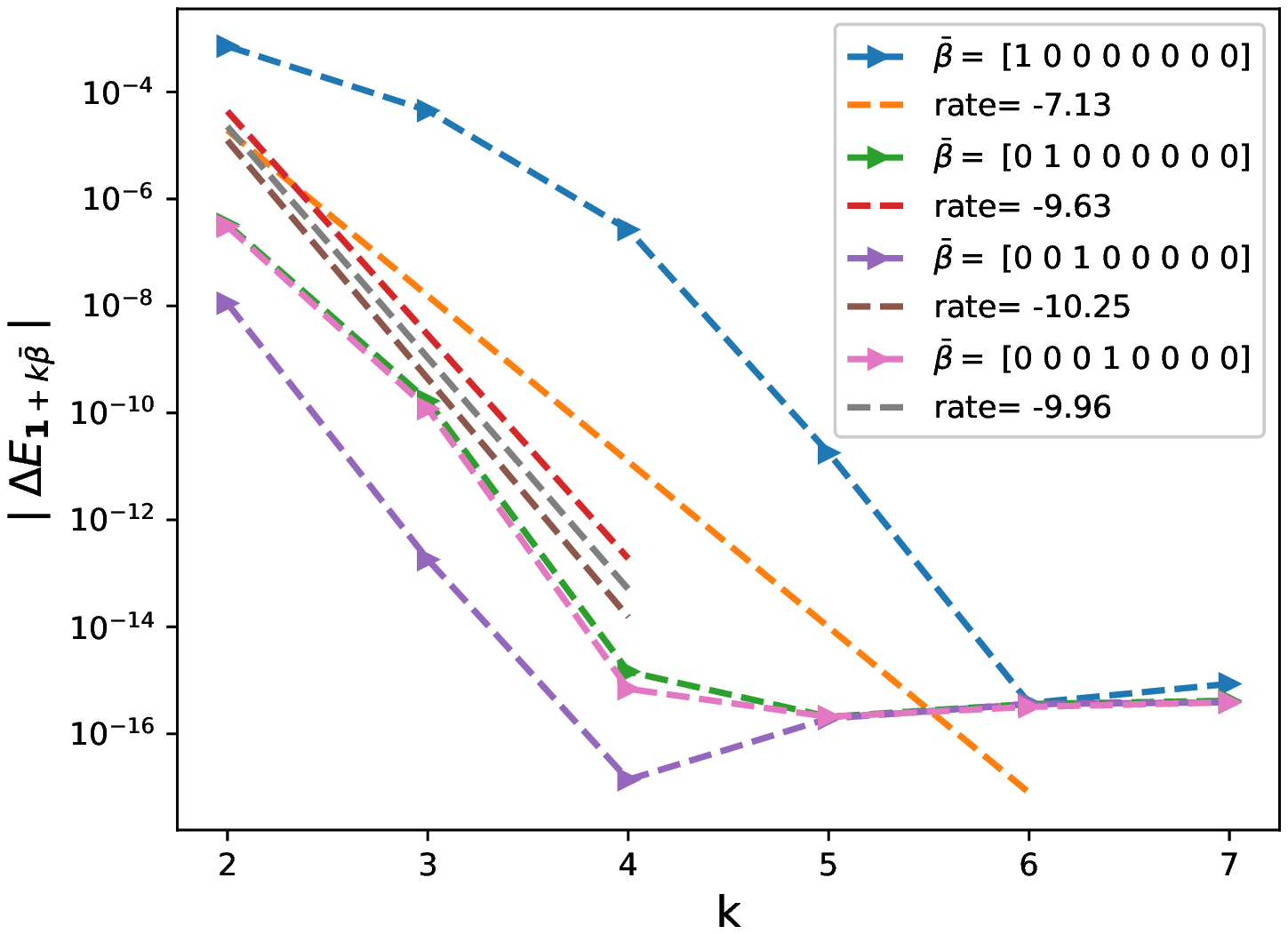}
		\caption{}
		\label{fig:sub3}
	\end{subfigure}%
	\begin{subfigure}{.4\textwidth}
		\centering
		\includegraphics[width=1\linewidth]{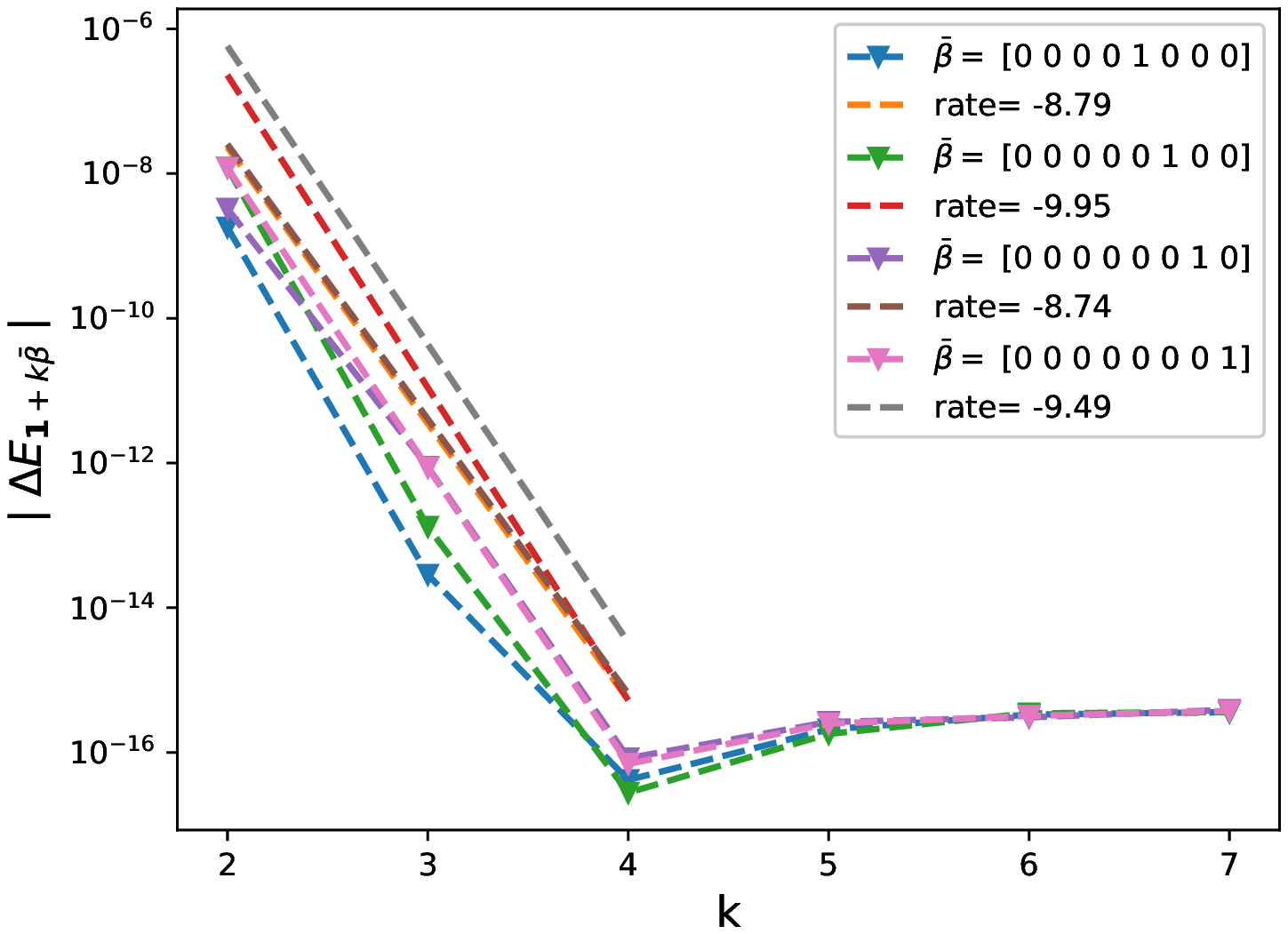}
		\caption{}
		\label{fig:sub4}
	\end{subfigure}

	\caption{The rate of error convergence of first order differences $\abs{\Delta \text{E}_{\boldsymbol{\beta}}}$, defined by \eqref{eq:Work_error_contributions}, ($\boldsymbol{\beta}=\mathbf{1}+k \bar{\boldsymbol{\beta}}$)  for parameter set $2$ in Table \ref{table:Reference solution, using MC with $500$ time steps, of Call option price under rBergomi model, for different parameter constellation.}. The number of quadrature points used in the $i$-th dimension is $N_i=2^{\beta_i-1}+1$ . a) With respect to $\mathbf{W}^{(1)}$. b) With respect to $\mathbf{W}^{(2)}$.}
	\label{fig:first_diff_comp_K_1_H_002}
\end{figure}

\begin{figure}[h!]
	\centering
	\begin{subfigure}{.4\textwidth}
		\centering
		\includegraphics[width=1\linewidth]{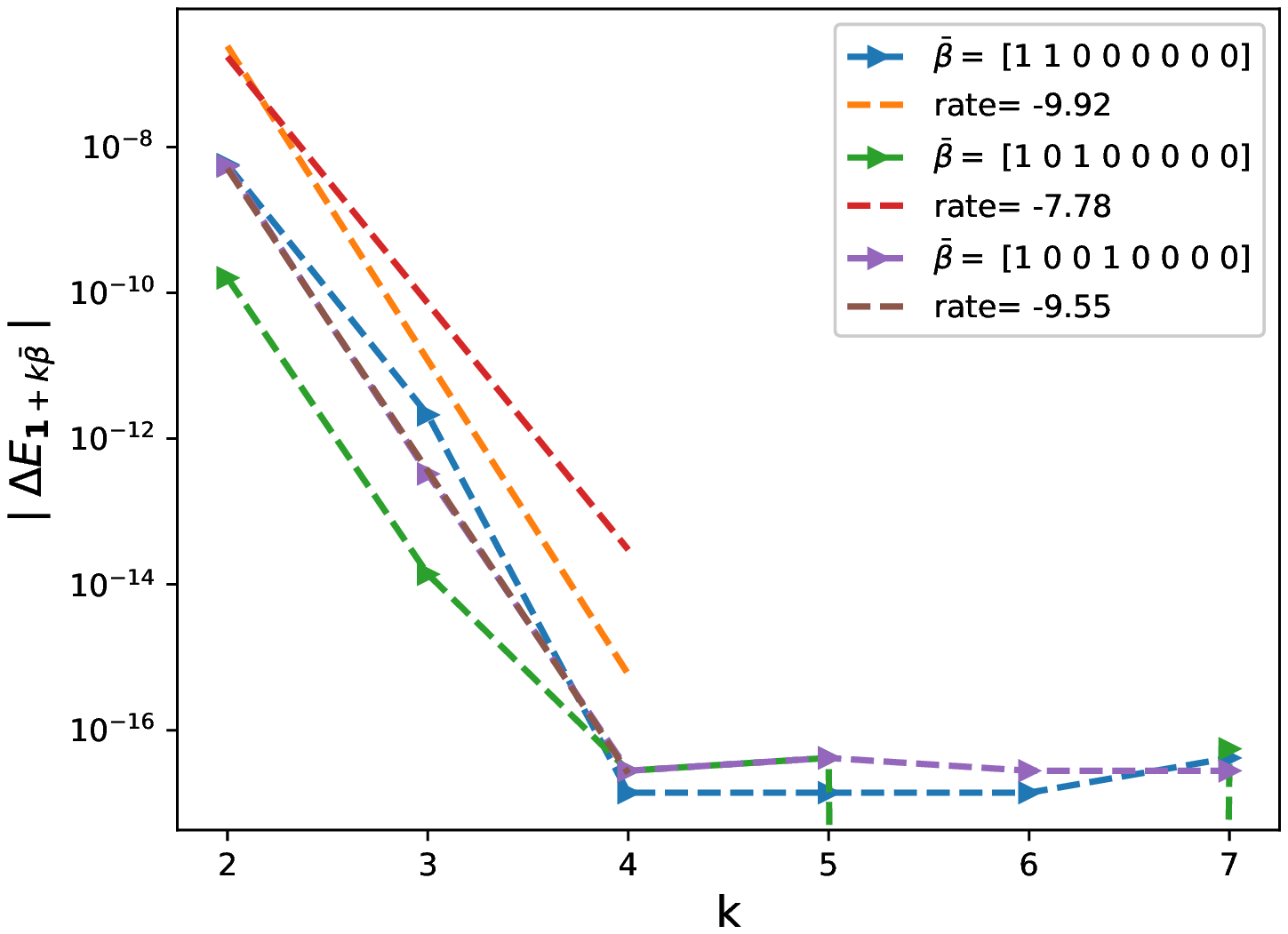}
		\caption{}
		\label{fig:sub3}
	\end{subfigure}%
	\begin{subfigure}{.4\textwidth}
		\centering
		\includegraphics[width=1\linewidth]{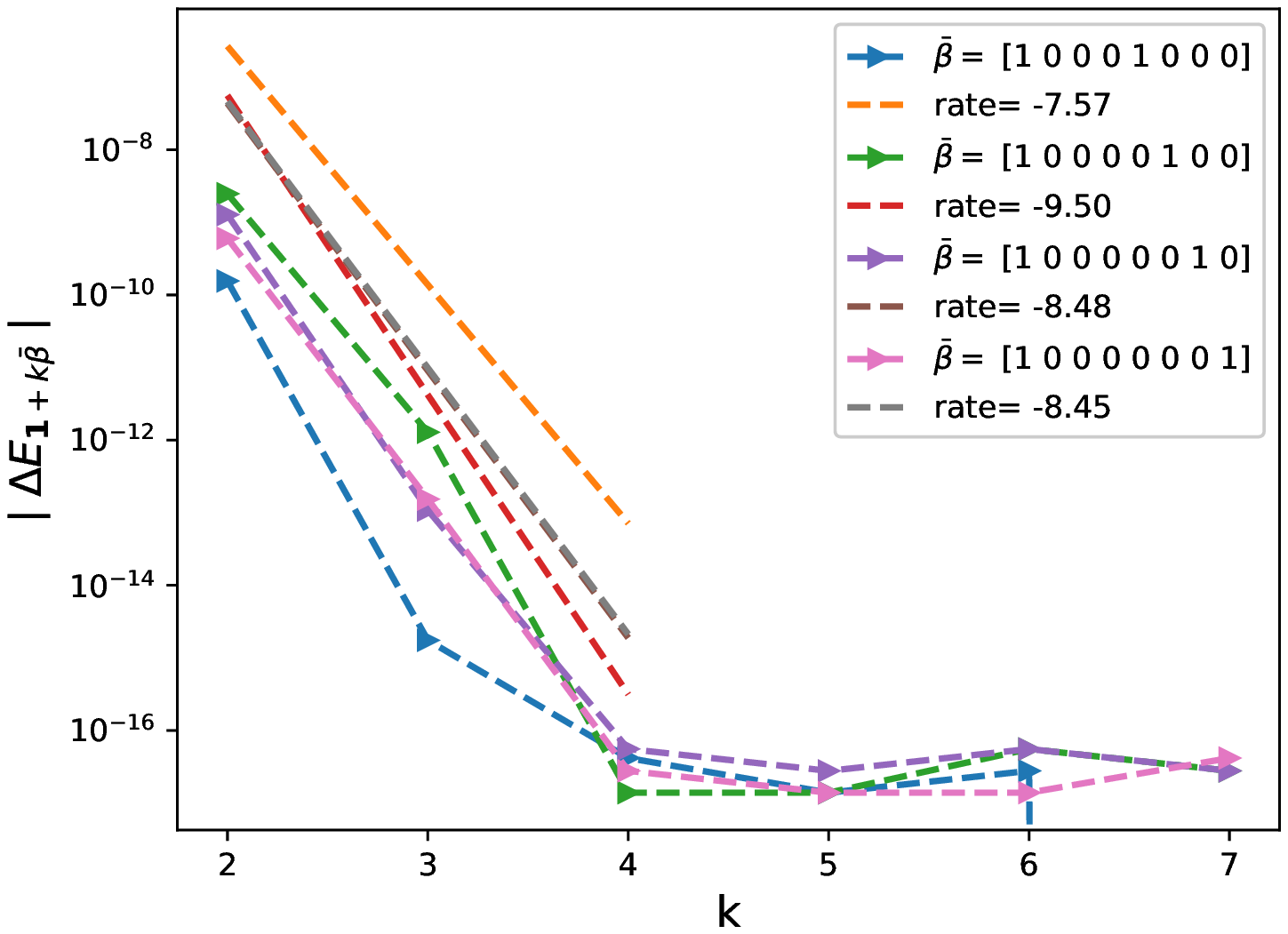}
		\caption{}
		\label{fig:sub4}
	\end{subfigure}
	
	\caption{The rate of error convergence of  second order differences $\abs{\Delta \text{E}_{\boldsymbol{\beta}}}$, defined by \eqref{eq:Work_error_contributions},  ($\boldsymbol{\beta}=\mathbf{1}+k \bar{\boldsymbol{\beta}}$) for parameter set $2$ in Table \ref{table:Reference solution, using MC with $500$ time steps, of Call option price under rBergomi model, for different parameter constellation.}. The number of quadrature points used in the $i$-th dimension is $N_i=2^{\beta_i-1}+1$. a) With respect to $\mathbf{W}^{(1)}$. b)  With respect to $\mathbf{W}^{(2)}$.}
	\label{fig:second_diff_comp_K_1_H_002}
\end{figure}

\begin{remark}
	The choice of the hierarchy of quadrature points, $m(\boldsymbol{\beta})$, is flexible in the ASGQ algorithm and can be fixed by the user, depending on the convergence properties of the problem at hand. For instance, for the sake of reproducibility, in our numerical experiments, we used a linear hierarchy: $m(\beta)=4 (\beta-1)+1,\: 1 \le \beta $, for results of parameter set $1$ in Table \ref{table:Reference solution, using MC with $500$ time steps, of Call option price under rBergomi model, for different parameter constellation.}. For the remaining parameter sets in Table  \ref{table:Reference solution, using MC with $500$ time steps, of Call option price under rBergomi model, for different parameter constellation.}, we used a geometric hierarchy: $m(\beta)=2^{\beta-1}+1, \:1 \le \beta $.
\end{remark} 
\begin{remark}
As emphasized in \cite{haji2016multi}, one important requirement to achieve the optimal performance of the ASGQ is to check  the error convergence, defined by \eqref{eq:Work_error_contributions},  of first and mixed difference operators. We checked this requirement in all our numerical experiments, and for illustration, we show in Figures  \ref{fig:first_diff_comp_K_1_H_002} and \ref{fig:second_diff_comp_K_1_H_002} the error convergence of first and second order differences for the case of parameter set $2$ in Table \ref{table:Reference solution, using MC with $500$ time steps, of Call option price under rBergomi model, for different parameter constellation.}.  These plots show that: i) $\Delta \text{E}_{\boldsymbol{\beta}}$ decreases exponentially fast with respect to $\beta_i$, and ii) $\Delta \text{E}_{\boldsymbol{\beta}}$ has a  product structure since  we  observe  a faster error decay for second differences, compared to corresponding first difference operators.
\end{remark} 

\FloatBarrier

\begin{remark}
The analiticity assumption, stated in the beginning of Section \ref{sec:Details of the MISC}, is crucial for the optimal performance of our proposed method. In fact, although we face the issue of the  ``curse of dimensionality" when increasing $N$, the analiticity of $f$ implies a spectral convergence for sparse grids quadrature.
\end{remark} 
\subsection{Quasi Monte Carlo (QMC)}\label{sec:Quasi Monte Carlo (QMC)}
A second type of deterministic quadrature that we test in this work is the randomized QMC method. Specifically, we use the lattice rules family of QMC \cite{sloan1985lattice,cools2008belgian,nuyens2014construction}.  The main input for the lattice rule is one integer vector with $d$ components ($d$ is the dimension of the integration problem).

In fact, given an integer vector $z = (z_1,\dots, z_d)$ known as \textit{the generating vector}, a (rank-$1$) lattice rule with $n$ points takes the form

\begin{equation}
Q_n(f):=\frac{1}{n}\sum_{k=0}^{n-1} f \left( \frac{kz \: \text{mod}\: n}{n}\right).
\end{equation}
The quality of the lattice rule depends on the choice of the generating vector. Due to the modulo operation, it is sufficient to consider the values from $1$ up to $n-1$. Furthermore, we restrict the values to those relatively prime to $n$, to ensure that every one-dimensional projection of the $n$ points yields $n$ distinct values. Thus, we write $\mathbf{z} \in \mathbb{U}_n^d$, with $\mathbb{U}_n:=\{z \in \zset: 1 \le z \le n-1\: \text{and gcd}(z,n)=1\}$. For practical purposes,  we choose $n$  to be a power of $2$. The total number of possible choices for the generating vector is then  $(n/2)^d$. 

To get an unbiased approximation of the integral, we use  a randomly shifted lattice rule, which also allows us to obtain a practical error estimate in the same way as the MC method. It works as follows. We generate $q$ independent random shifts $\Delta^{(i)}$ for $i=0,\dots,q-1$ from the uniform distribution of $[0,1]^d$ . For the same fixed lattice generating vector $z$, we compute the $q$ different shifted lattice rule approximations and denote
them by $Q^{(i )}_n(f)$ for $i=0,\dots,q-1$.We then take the average
\begin{align}
\overline{Q}_{n,q}(f)=\frac{1}{q} \sum_{i=0}^{q-1}Q^{(i )}_n(f)=\frac{1}{q}\sum_{i=0}^{q-1}\left(\frac{1}{n}\sum_{k=0}^{n-1} f \left( \frac{kz+\Delta^{(i)}  \: \text{mod}\: n}{n}\right)  \right)
\end{align}
as our final approximation to the integral and the total number of samples of the randomized QMC method is $M^{\text{QMC}}=q \times n$.

We note that, since we are dealing with Gaussian randomness and with integrals in infinite support, we use the inverse of the standard normal cumulative distribution function as a pre-transformation to map the problem to $[0,1]$, and then use the randomized QMC. Furthermore, in our numerical test, we use a pre-made point generator using latticeseq\_b2.py in python from   \url{https://people.cs.kuleuven.be/~dirk.nuyens/qmc-generators/}.

\begin{remark}
We emphasize that we opt for one of the  most simple rules (very easy to specify and implement) for QMC, that is the lattice rule, which, thanks to the hierarchical representations that we implemented (Richardson extrapolation and Brownian bridge construction) appears to give already a very good performance. Furthermore, we believe that using more sophisticated rules such as  lattice sequences or interlaced Sobol sequences may have a similar or even better performance.
\end{remark}

\subsection{Brownian bridge  construction}\label{sec:Brwonian bridge construction}
In the literature on ASGQ and  QMC, several hierarchical path generation methods (PGMs) have been proposed to reduce the effective dimension. Among these techniques, we mention Brownian
bridge  construction \cite{morokoff1994quasi,moskowitz1996smoothness,caflisch1997valuation}, principal component analysis (PCA)  \cite{acworth1998comparison} and  linear transformation (LT) \cite{imai2004minimizing}.

In our context, the Brownian motion on a time discretization  can be constructed either sequentially using a standard random walk construction, or hierarchically using   other PGMs, as listed above. For our purposes, to make an effective use of ASGQ or QMC methods, which benefit from anisotropy, we use the Brownian
bridge construction that produces  dimensions that are not equally important, contrary to a random walk procedure for which all the dimensions of the stochastic space have equal importance. In fact, Brownian
bridge uses the first several coordinates of the low-discrepancy points to determine the general shape of the Brownian path, and the last few coordinates influence only the fine detail of the path. Consequently, this representation  reduces the effective dimension  of the problem, resulting in an acceleration of the ASGQ and QMC methods by reducing the computational cost.

Let us denote $\{t_i\}_{i=0}^{N}$ as the grid of time steps. Then the Brownian
bridge construction \cite{glasserman2004monte} consists of the following: given a past value $B_{t_i}$ and a future value $B_{t_k}$, the value $B_{t_j}$ (with $t_i < t_j < t_k$) can be generated according to 
\begin{equation*}
B_{t_j}=(1-\rho) B_{t_i}+\rho B_{t_k}+ \sqrt{\rho (1-\rho)(k-i) \Delta t} z, \: z \sim \mathcal{N}(0,1) \COMMA
\end{equation*}
where $\rho=\frac{j-i}{k-i}$.  

\subsection{Richardson extrapolation}\label{sec:Richardson extrapolation}
Another representation that we couple with the ASGQ and QMC methods is the Richardson extrapolation \cite{talay1990expansion}. In fact, applying level $K_\text{R}$ (level of extrapolation) of the Richardson extrapolation  dramatically reduces the bias, and as a consequence, reduces the  number of time steps $N$ needed in the coarsest level to achieve a certain error tolerance. As a consequence, the Richardson extrapolation directly reduces  the total dimension of the integration problem for achieving some error tolerance.

\begin{conjecture}\label{conj: Richardson extrapol}
Let us denote by $(X_t)_{0 \le t \le T}$  a stochastic process following the rBergomi dynamics given by \eqref{eq:rBergomi_model1},  and by $(\hat{X}_{t_i}^{\Delta t})_{0 \le  t_i \le T}$ its approximation using the hybrid  scheme (described in Section \ref{sec: The hybrid scheme}) with a time step $\Delta t$.  Then, for sufficiently small $\Delta t$, and a suitable smooth function $f$, we assume that
\begin{align}\label{Euler_weak_error_strenghten}
	\expt{f(\hat{X}_T^{\Delta t})}= \expt{f(X_T)} + c \Delta t +\Ordo{\Delta t^2} \PERIOD
\end{align}
\end{conjecture}
Applying \eqref{Euler_weak_error_strenghten} with discretization step $2 \Delta t$, we  obtain
\begin{align*}
	\expt{f(\hat{X}_T^{2 \Delta t})}= \expt{f(X_T)} + 2 c \Delta t +\Ordo{\Delta t^2} \COMMA
\end{align*}
implying
\begin{align*}
	2 \expt{f(\hat{X}_T^{2 \Delta t})}- \expt{f(\hat{X}_T^{\Delta t})} =\expt{f(X_T)} + \Ordo{\Delta t^2} \PERIOD
\end{align*}
For higher levels of extrapolations, we use the following: Let us denote by $\Delta t_J=\Delta t_0 2^{-J}$ the grid sizes (where $\Delta t_0$ is the coarsest grid size), by $K_\text{R}$ the level of the Richardson extrapolation, and by $I(J,K_\text{R})$ the approximation of $\expt{f(X_T)}$ by terms up to level $K_\text{R}$ (leading to a weak error of order $K_\text{R}$), then we have the following recursion 
\begin{align*}
I(J,K_\text{R})=\frac{2^{K_\text{R}}I(J,K_\text{R}-1)-I(J-1,K_\text{R}-1)}{2^{K_\text{R}}-1},\quad J=1,2,\dots, K_\text{R}=1,2,\dots
\end{align*}
\begin{remark}
We emphasize that, throughout our work, we are interested in the pre-asymptotic regime (a small number of time steps), and the use of Richardson extrapolation is justified by conjectures  \ref{conj: Weak error structure} and \ref{conj: Richardson extrapol},  and our observed experimental results in that regime (see Section \ref{sec:Weak error plots_no_change}),  which suggest  a convergence of order one for the weak error. 
\end{remark}

 \section{Numerical experiments}\label{sec:Numerical tests}
In this section, we show the results obtained through the different numerical experiments,  conducted across different parameter constellations for the rBergomi model. Details about these examples are presented in Table \ref{table:Reference solution, using MC with $500$ time steps, of Call option price under rBergomi model, for different parameter constellation.}. The first set is the one  closest to the empirical findings \cite{bennedsen2016decoupling,gatheral2018volatility}, which suggest that $H \approx 0.1$. The choice of parameter values of $\nu= 1.9$ and $\rho=-0.9$ is justified by \cite{bayer2016pricing}, where it is shown that these values are remarkably consistent with the SPX market on $4$th February $2010$. For the remaining three sets in Table \ref{table:Reference solution, using MC with $500$ time steps, of Call option price under rBergomi model, for different parameter constellation.}, we want to test the potential of our method for a very rough case, that is $H=0.02$, for three different  scenarios  of moneyness, $S_0/K$. In fact, hierarchical variance reduction methods, such as MLMC, are inefficient in this context because of the poor behavior of the strong error,    which is of the order of $H$ \cite{neuenkirch2016order}. We emphasize that we checked the robustness of our method for other parameter sets, but for illustrative purposes, we only show results for the parameters sets presented in Table \ref{table:Reference solution, using MC with $500$ time steps, of Call option price under rBergomi model, for different parameter constellation.}. For all our numerical experiments, we consider   a number of time steps $N \in \{2,4,8,16\}$, and  all reported errors are relative errors, normalized by the reference solutions provided in Table \ref{table:Reference solution, using MC with $500$ time steps, of Call option price under rBergomi model, for different parameter constellation.}.

\FloatBarrier
\begin{table}[!h]
	\centering
	\begin{small}
	\begin{tabular}{l*{2}{c}r}
	\toprule[1.5pt]
		Parameters            & Reference solution    \\
		\hline

			Set $1$:	$H=0.07, K=1,S_0=1, T=1, \rho=-0.9, \eta=1.9,\xi_0=0.235^2$   & $\underset{(5.6e-05)}{0.0791}$  \\	

				Set $2$:	$H=0.02, K=1, S_0=1, T=1,\rho=-0.7, \eta=0.4,\xi_0=0.1$   & $\underset{(9.0e-05)}{0.1246}$  \\
					Set $3$:	$H=0.02, K=0.8,S_0=1,T=1, \rho=-0.7, \eta=0.4,\xi_0=0.1$   & $\underset{(5.4e-05)}{0.2412}$  \\
						Set $4$:	$H=0.02, K=1.2,S_0=1,T=1, \rho=-0.7, \eta=0.4,\xi_0=0.1$   & $\underset{(8.0e-05)}{0.0570}$  \\
	\bottomrule[1.25pt]
	\end{tabular}
\end{small}
	\caption{Reference solution, which is the  approximation of the call option price under the rBergomi model, defined in \eqref{BS_formula_rbergomi},  using MC with $500$ time steps and number of samples, $M=8 \times 10^6$, for different parameter constellations.  The numbers between parentheses correspond to the statistical error estimates.}
	\label{table:Reference solution, using MC with $500$ time steps, of Call option price under rBergomi model, for different parameter constellation.}
\end{table}
\FloatBarrier

\subsection{Weak error} \label{sec:Weak error plots_no_change}
\begin{figure}[h!]
	\centering
	\begin{subfigure}{.4\textwidth}
		\centering
		\includegraphics[width=1\linewidth]{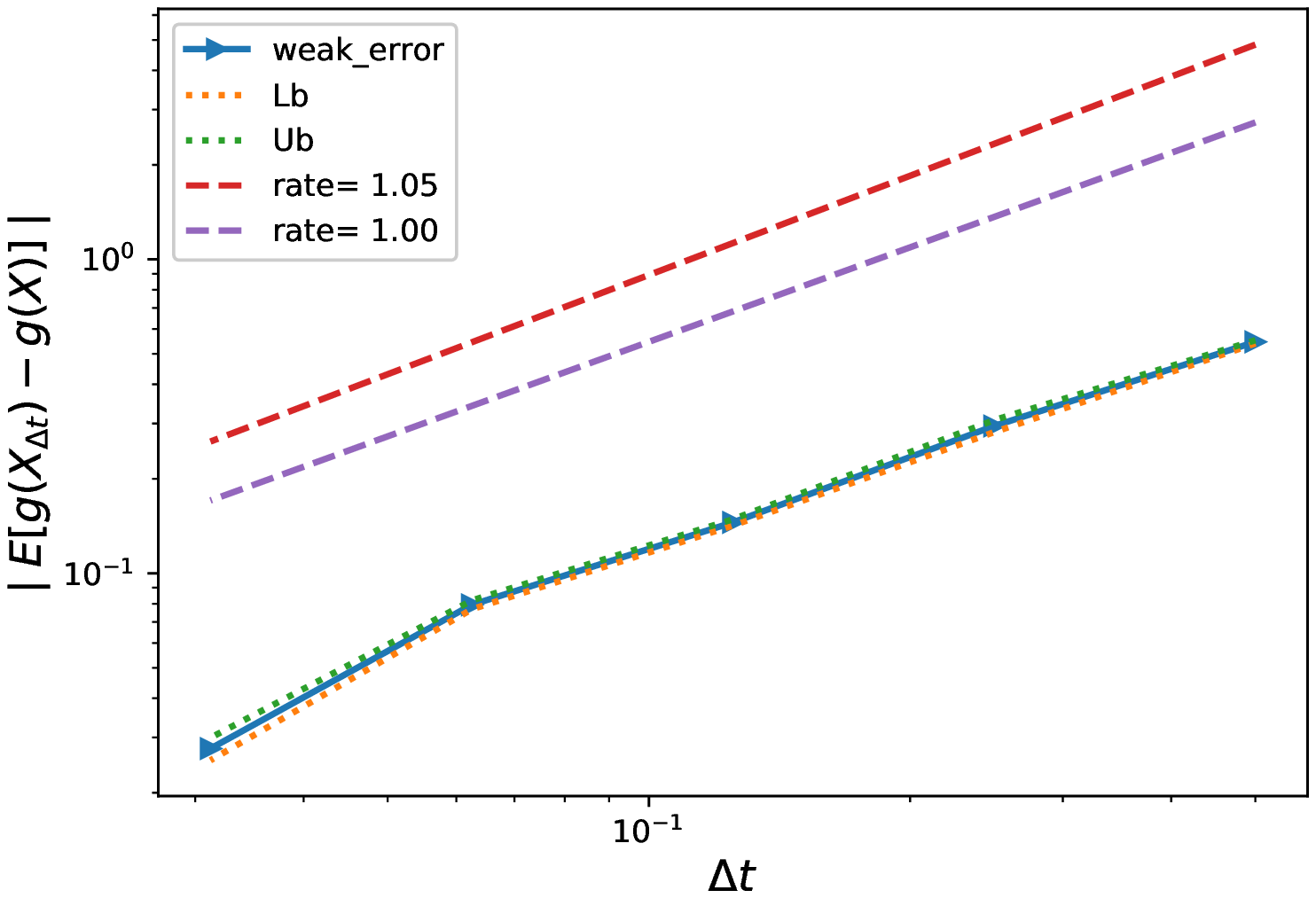}
		\caption{}
		\label{fig:sub3}
	\end{subfigure}%
	\begin{subfigure}{.4\textwidth}
		\centering
		\includegraphics[width=1\linewidth]{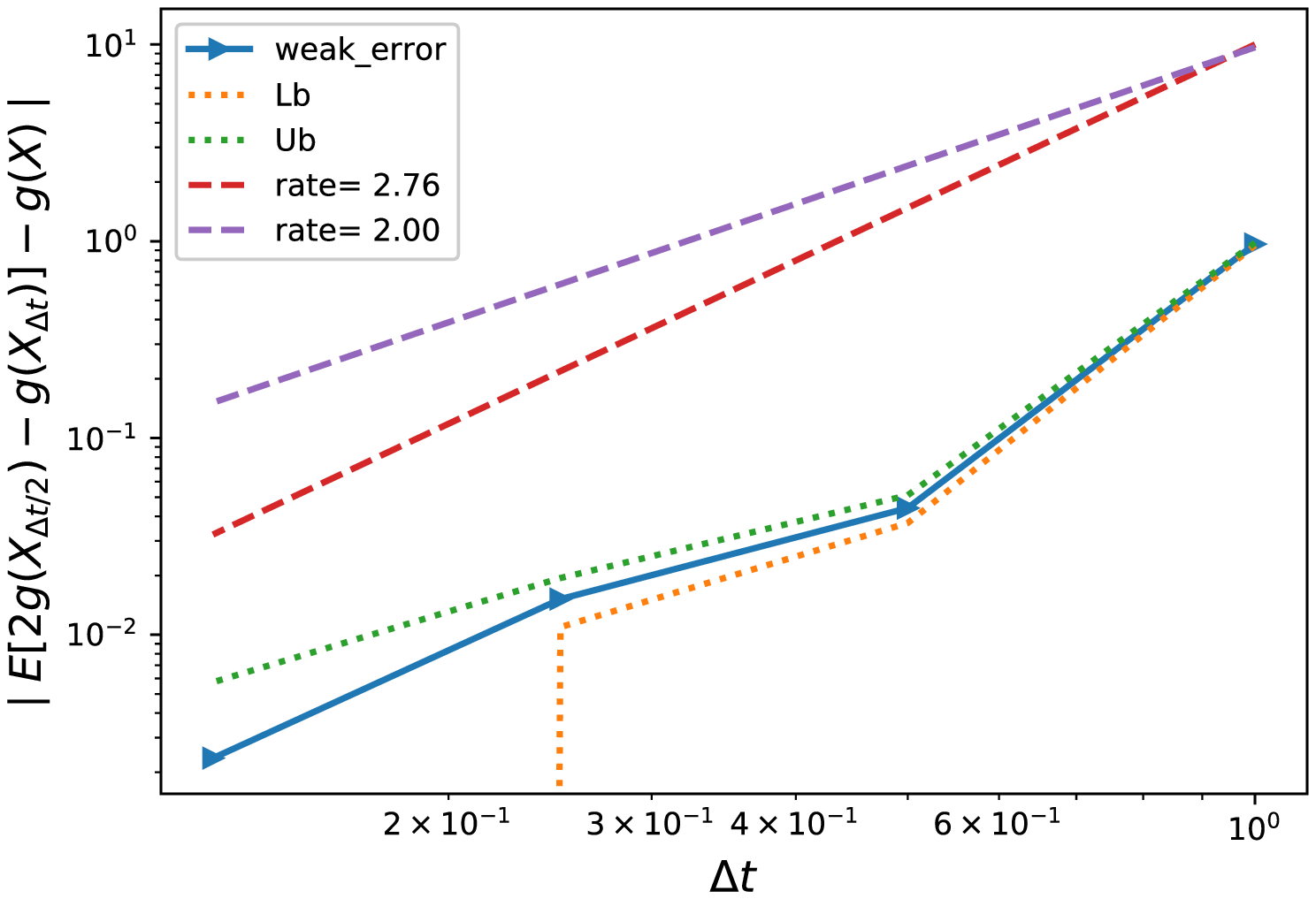}
		\caption{}
		\label{fig:sub4}
	\end{subfigure}
	
	\caption{The  convergence of the weak error $\mathcal{E}_B(N)$, defined in \eqref{eq: Weak_error_hyb_chol}, using MC, for set $1$ parameter in Table \ref{table:Reference solution, using MC with $500$ time steps, of Call option price under rBergomi model, for different parameter constellation.}. We refer to $C_{\text{RB}}$ as $\expt{g(X)}$, and to $C_{\text{RB}}^{N}$ as  $\expt{g(X_{\Delta t})}$. The upper and lower bounds are $95\%$ confidence intervals. a) without Richardson extrapolation.  b) with Richardson extrapolation (level $1$).}
	\label{fig:Weak_rate_set1_set_2_without_rich}
\end{figure}

We start our numerical experiments by accurately  estimating the weak error (bias) discussed in Section \ref{sec:Weak error analysis}, for the different  sets of parameters in Table \ref{table:Reference solution, using MC with $500$ time steps, of Call option price under rBergomi model, for different parameter constellation.}, with and without the Richardson extrapolation.   For illustrative purposes, we only show the weak errors related to set $1$ in Table \ref{table:Reference solution, using MC with $500$ time steps, of Call option price under rBergomi model, for different parameter constellation.} (see Figure \ref{fig:Weak_rate_set1_set_2_without_rich}). We note that we observed a similar behavior for the other parameter sets, with slightly worse rates for some cases. We emphasize that the reported weak rates correspond to the pre-asymptotic regime that we are interested in. We are not interested in estimating the rates specifically but rather obtaining  a sufficiently precise estimate of the weak error (bias), $\mathcal{E}_B(N)$, for different  numbers of time steps $N$.  For a fixed discretization, the corresponding estimated biased solution will be set as a reference solution to the  ASGQ method  in order to estimate the quadrature error $\mathcal{E}_Q(\text{TOL}_{\text{ASGQ}},N)$.	
\FloatBarrier
\FloatBarrier
\subsection{Comparing the errors and computational time for MC, QMC and ASGQ}\label{sec:Comparing different  errors and complexity for MC and MISC}
In this section, we conduct a comparison between MC, QMC and ASGQ, in terms of errors and computational time. We show tables and plots reporting  the different relative errors involved in the MC and QMC  methods (bias and statistical error  estimates), and in ASGQ (bias and quadrature error estimates).  While fixing  a  sufficiently small relative error tolerance in the price estimates,  we  compare the computational time needed for all methods to meet the desired error tolerance.  We note that,  in all cases, the actual work (runtime) is obtained using an Intel(R) Xeon(R) CPU E$5$-$268$ architecture. 

For the numerical experiments we conducted for each parameter set, we follow these steps to achieve the reported results:
\begin{enumerate}
\item[i)] For a fixed number of time steps, $N$, we compute an accurate estimate, using a large number of samples, $M$, of the biased  MC solution, $C_{RB}^{N}$. This step also provides us with an estimate of the bias error, $\mathcal{E}_B(N)$, defined by \eqref{eq:total_error_ASGQ}. 
\item[ii)] The estimated  biased solution,  $C_{\text{RB}}^{N}$, is used as a reference solution  to the ASGQ method to compute the quadrature error, $\mathcal{E}_Q(\text{TOL}_{\text{ASGQ}},N)$, defined by \eqref{eq:quadrature error}.
\item[iii)] In order to compare the different methods, the number of samples, $M^{\text{QMC}}$ and $M^{\text{MC}}$, are chosen so that  the statistical errors of randomized QMC, $\mathcal{E}_{S,\text{QMC}}(M^{\text{QMC}})$, and MC, $\mathcal{E}_{S,\text{MC}}(M^{\text{MC}})$, satisfy
\begin{align}\label{optimal_number_samples}
\mathcal{E}_{S,\text{QMC}}(M^{\text{QMC}})=\mathcal{E}_{S,\text{MC}}(M^{\text{MC}})= \mathcal{E}_B(N)=\frac{\mathcal{E}_{\text{tot}}}{2}\COMMA
\end{align}
where $\mathcal{E}_B(N)$ is the bias as defined in \eqref{eq:total_error_ASGQ} and
$\mathcal{E}_{\text{tot}}$ is the total error. 
\end{enumerate}

We show  the summary of our numerical findings in Table \ref{table:Summary of our numerical results.}, which  highlights the computational gains achieved by ASGQ and QMC over the MC method to meet a certain error tolerance, which we set approximately to $1\%$. We note that the results are reported using the best configuration with Richardson extrapolation for each method. More detailed results for each case of parameter set, as in Table \ref{table:Reference solution, using MC with $500$ time steps, of Call option price under rBergomi model, for different parameter constellation.},  are provided in  Sections \ref{sec:Case of set $2$ parameters_linear}, \ref{sec:Case of set 3 parameters}, \ref{sec:Case of set 4 parameters} and \ref{sec:Case of set 5 parameters}. 
\FloatBarrier
\begin{table}[!h]
	\centering
	\begin{small}
	\begin{tabular}{l*{4}{c}r}
	\toprule[1.5pt]
		Parameter set              &  Total relative error  & CPU time ratio $\left(\text{ASGQ}/\text{MC} \right)$ & CPU time ratio  $\left( \text{QMC}/\text{MC} \right)$\\
		\hline
			Set $1$  &  $1\%$&  $ 6.7\%$ &  $10\%$\\	
           	
              \hline
            Set $2$     &  $0.2\%$&  $4.7\%$ &  $1.4\%$\\		
				 \hline
					Set $3$    &  $0.4\%$&  $3.8\%$ &  $4.7\%$\\	
					\hline
						Set $4$  &  $2\%$&  $20\%$ &  $10\%$\\	
		\bottomrule[1.25pt]
	\end{tabular}
\end{small}
	\caption{Summary of relative errors and computational gains, achieved by the different methods. In this table, we highlight the computational gains achieved by ASGQ and QMC over the MC method to meet a certain error tolerance. We note that the ratios are computed for the best configuration with the Richardson extrapolation for each method. We provide details about how we compute these gains, for each case, in Sections \ref{sec:Case of set $2$ parameters_linear}, \ref{sec:Case of set 3 parameters}, \ref{sec:Case of set 4 parameters}, and \ref{sec:Case of set 5 parameters}.}
	\label{table:Summary of our numerical results.}
\end{table}
\FloatBarrier

\subsubsection{Case of parameters in Set 1 in Table \ref{table:Reference solution, using MC with $500$ time steps, of Call option price under rBergomi model, for different parameter constellation.} }
\label{sec:Case of set $2$ parameters_linear}
\begin{figure}[htb]
	\centering 
	\begin{subfigure}{0.33\textwidth}
		\includegraphics[width=\linewidth]{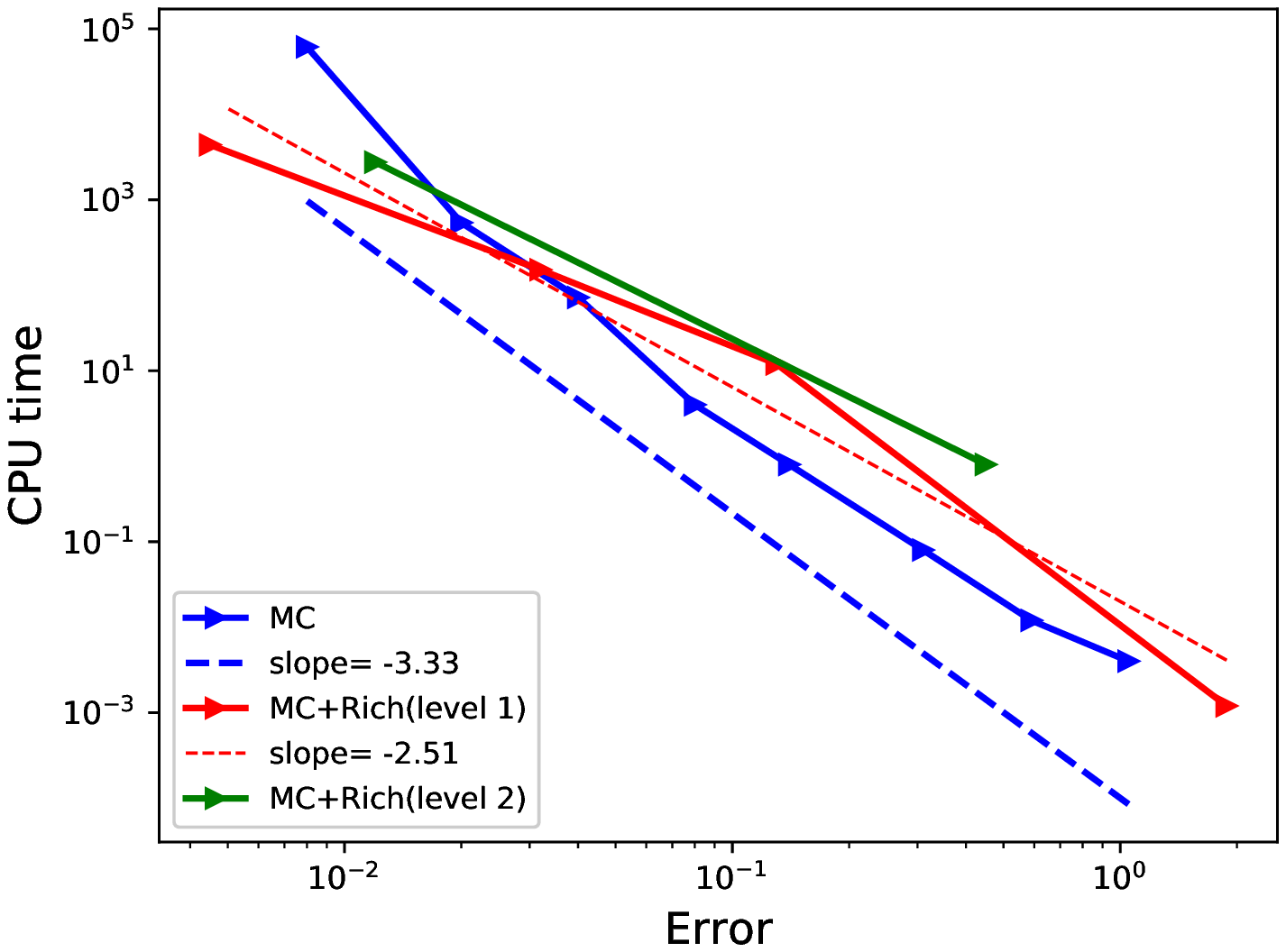}
		\caption{}
		\label{fig:1}
	\end{subfigure}\hfil 
	\begin{subfigure}{0.33\textwidth}
		\includegraphics[width=\linewidth]{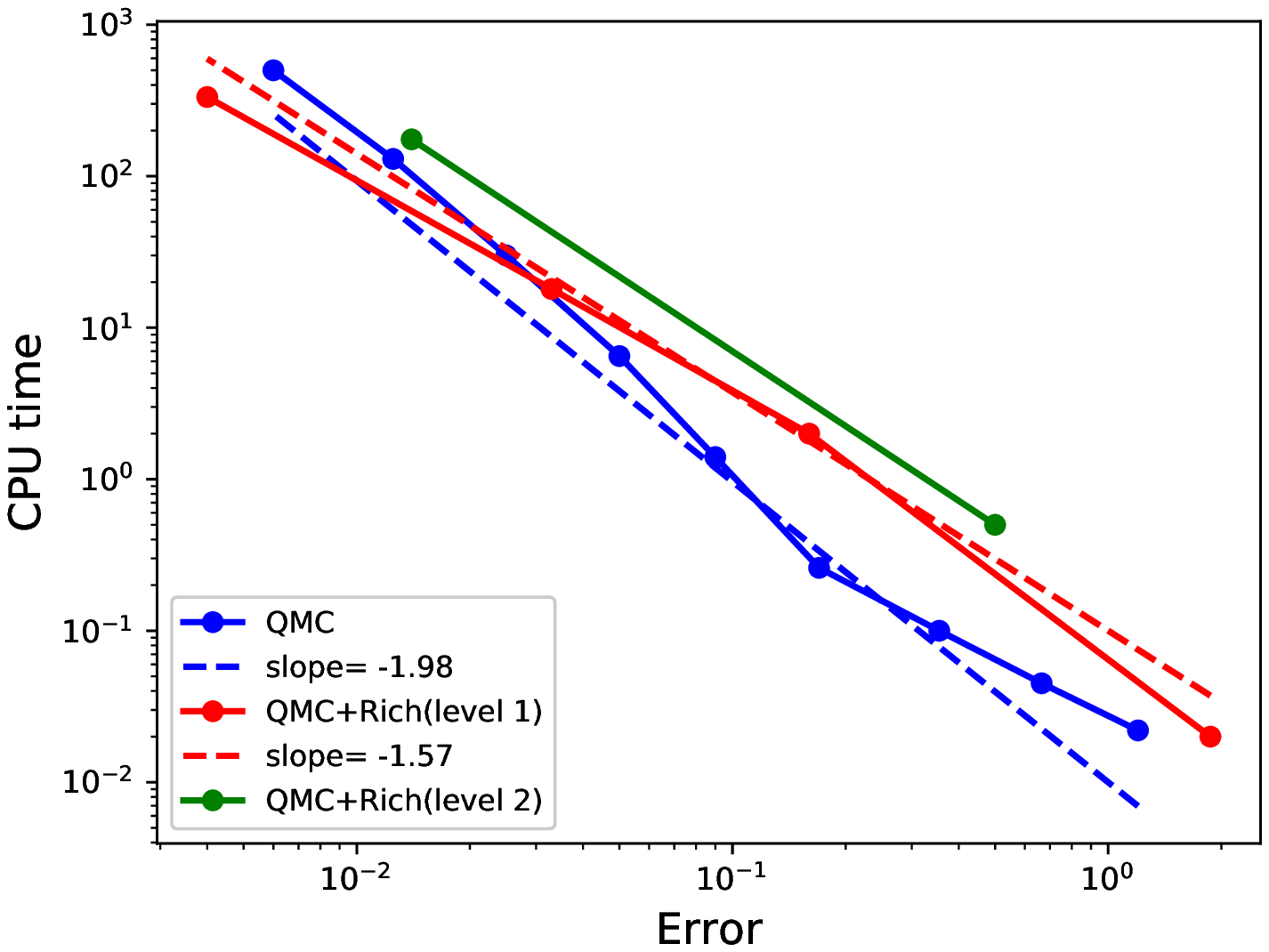}
		\caption{}
		\label{fig:2}
	\end{subfigure}\hfil 
	\begin{subfigure}{0.33\textwidth}
		\includegraphics[width=\linewidth]{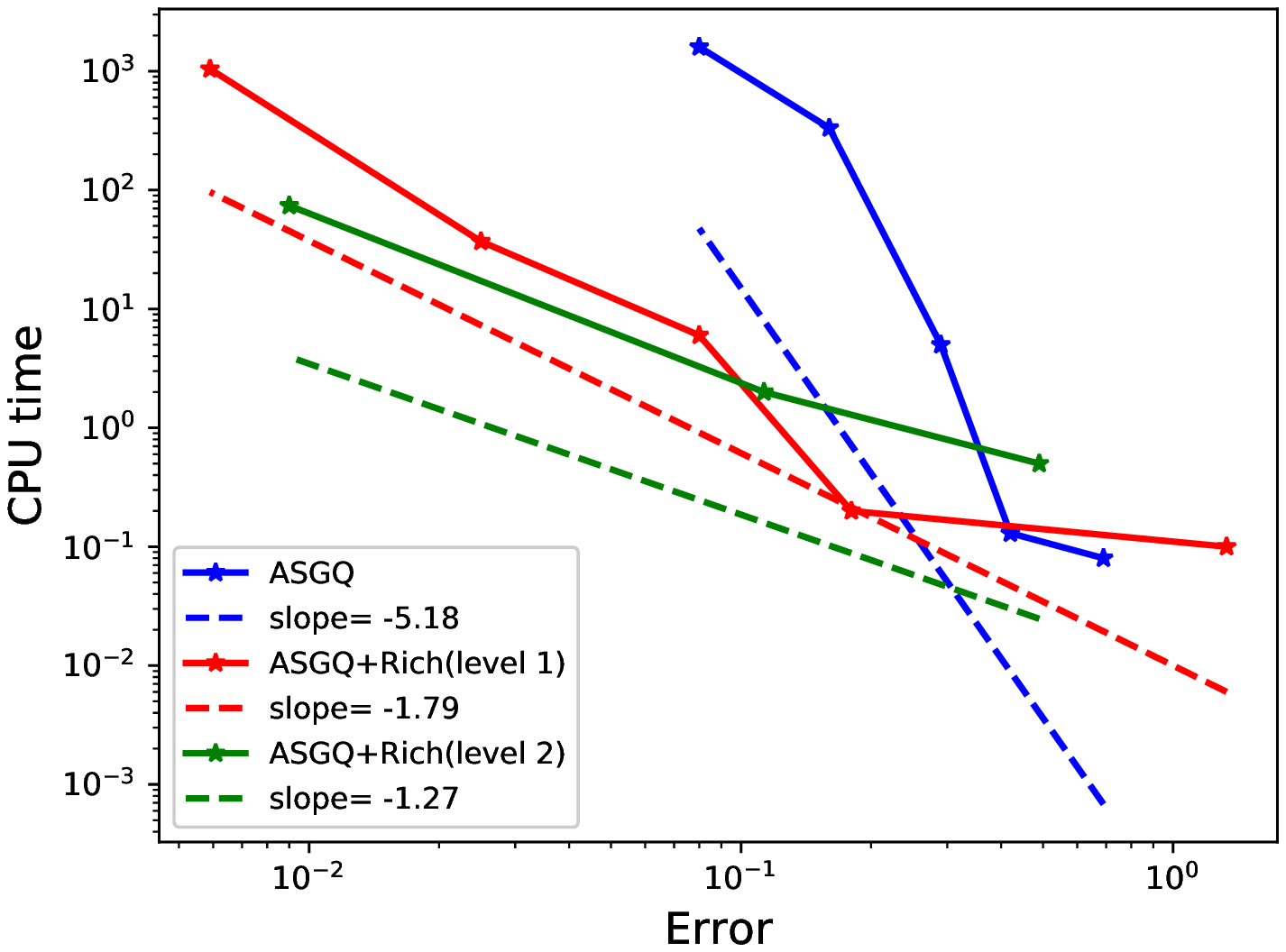}
		\caption{}
		\label{fig:3}
	\end{subfigure}
	\caption{Comparing the numerical complexity of the different  methods with the different configurations in terms of the level of Richardson extrapolation, for the case of parameter set $1$ in Table \ref{table:Reference solution, using MC with $500$ time steps, of Call option price under rBergomi model, for different parameter constellation.}. a) MC methods. b) QMC methods. d) ASGQ methods.}
	\label{fig: Comparing the numerical complexity of the different  methods with the different configurations}
\end{figure}

In this section, we conduct our numerical experiments for three different scenarios: i) without Richardson extrapolation, ii) with (level $1$) Richardson extrapolation, and iii) with (level $2$) Richardson extrapolation. Figure \ref{fig: Comparing the numerical complexity of the different  methods with the different configurations}  shows a comparison of the numerical complexity for each method under the three different scenarios. From this Figure, we conclude that, to achieve a relative error of $1\%$, the level $1$ of the Richardson extrapolation is the optimal configuration for both the MC and the randomized QMC methods, and  that the level $2$ of the Richardson extrapolation is the optimal configuration for the ASGQ  method.

\FloatBarrier
\FloatBarrier
We compare these optimal configurations for each method  in Figure \ref{fig:Complexity plot for  MISC for Case set $2$ parameters, comparison}, and we show that both ASGQ and QMC outperform MC, in terms of numerical complexity. In particular, to achieve a total relative error of $1\%$, ASGQ coupled with the level $2$ of the Richardson extrapolation requires
approximately $6.7\%$ of the work of MC coupled with the level $1$ of the Richardson extrapolation, and  QMC coupled with the level $1$ of the Richardson extrapolation requires approximately $10\%$ of the work of MC coupled with the level $1$ of the Richardson extrapolation. We show more detailed outputs for the methods that are compared in Figure \ref{fig:Complexity plot for  MISC for Case set $2$ parameters, comparison} in Appendix \ref{appendix:Case of set 1 parameters}.
\FloatBarrier
\begin{figure}[h!]
	\centering
	\includegraphics[width=0.4\linewidth]{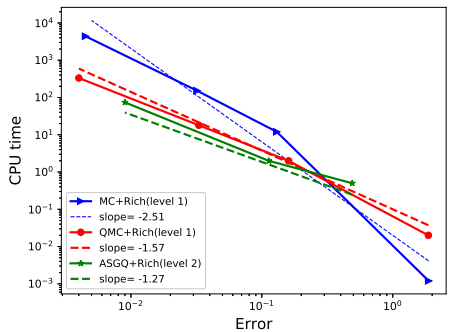}
	
	\caption{Computational work comparison for the different methods with the best configurations, as concluded from Figure \ref{fig: Comparing the numerical complexity of the different  methods with the different configurations}, for the case of parameter set $1$ in Table \ref{table:Reference solution, using MC with $500$ time steps, of Call option price under rBergomi model, for different parameter constellation.}. This plot shows that, to achieve a relative error below $1\%$, ASGQ coupled with the level $2$ of the Richardson extrapolation and QMC coupled with the level $1$ of the  Richardson extrapolation have a similar performance. Furthermore, they significantly outperform  the MC method coupled with the level $1$ of the Richardson extrapolation.}
	\label{fig:Complexity plot for  MISC for Case set $2$ parameters, comparison}
\end{figure}
\FloatBarrier

\subsubsection{Case of parameters in Set 2  in Table \ref{table:Reference solution, using MC with $500$ time steps, of Call option price under rBergomi model, for different parameter constellation.} }\label{sec:Case of set 3 parameters}

In this section, we only conduct our numerical experiments for the case without the Richardson extrapolation, since the results show that we meet a small enough relative error tolerance without the need to apply the Richardson extrapolation. We compare the different methods  in Figure \ref{fig:Complexity plot for  MISC for case set $3$ parameters, comparison}, and we determine that both ASGQ and QMC outperform MC, in terms of numerical complexity. In particular,  to achieve a total relative error of approximately $0.2\%$, ASGQ  requires
	approximately $4.7\%$ of the work of MC, and that QMC requires approximately $1.4\%$ of the work of MC. We show more detailed outputs for the methods that are compared in Figure \ref{fig:Complexity plot for  MISC for case set $3$ parameters, comparison} in Appendix \ref{appendix:Case of set $2$ parameters}. 
\FloatBarrier
	\begin{figure}[h!]
	\centering
	\includegraphics[width=0.4\linewidth]{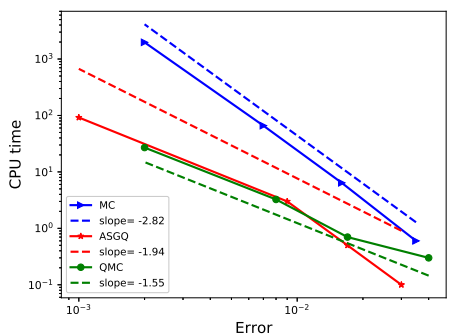}
	
	\caption{Computational work comparison for   the different  methods, for the case of parameter set $2$ in Table \ref{table:Reference solution, using MC with $500$ time steps, of Call option price under rBergomi model, for different parameter constellation.}. This plot shows that, to achieve a relative error below $1\%$, ASGQ and QMC have a similar performance and that they outperform  the MC method significantly, in terms of computational time.}
	\label{fig:Complexity plot for  MISC for case set $3$ parameters, comparison}
\end{figure}
\FloatBarrier
\subsubsection{Case of parameters in Set 3  in Table \ref{table:Reference solution, using MC with $500$ time steps, of Call option price under rBergomi model, for different parameter constellation.} }\label{sec:Case of set 4 parameters}
\begin{figure}[h!]
	\centering
	\includegraphics[width=0.4\linewidth]{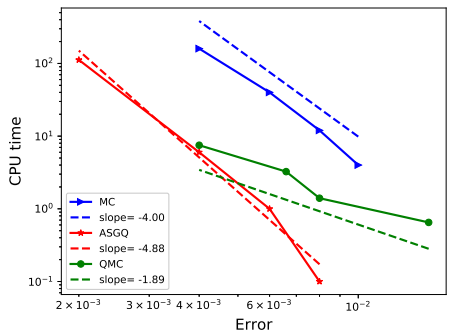}
	
	\caption{Comparison of computational work for the different  methods, for the case of parameter set $3$ in Table \ref{table:Reference solution, using MC with $500$ time steps, of Call option price under rBergomi model, for different parameter constellation.}.}
	\label{fig:Complexity plot for MC and MISC for case set $4$ parameters}
\end{figure}

In this section, we only conduct our numerical experiments for the case without the Richardson extrapolation, since the results show that we meet a small enough relative error tolerance without the need to apply the Richardson extrapolation. We compare the different methods  in Figure \ref{fig:Complexity plot for MC and MISC for case set $4$ parameters}, and we determine that both ASGQ and QMC outperform MC, in terms of numerical complexity. In particular,  to achieve a total relative error of approximately $0.4\%$, ASGQ  requires
	approximately $3.8\%$ of the work of MC, and that QMC requires approximately $4.7\%$ of the work of MC. We show more detailed outputs for the methods that are compared in Figure \ref{fig:Complexity plot for MC and MISC for case set $4$ parameters} in Appendix \ref{appendix:Case of set 3 parameters}.

\FloatBarrier

\FloatBarrier

\subsubsection{Case of parameters in Set 4  in Table \ref{table:Reference solution, using MC with $500$ time steps, of Call option price under rBergomi model, for different parameter constellation.} }\label{sec:Case of set 5 parameters}

\begin{figure}[h!]
	\centering
	\includegraphics[width=0.4\linewidth]{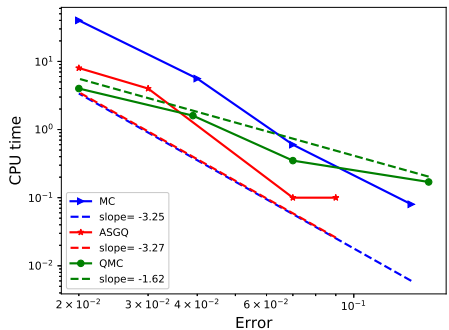}
	
	\caption{Comparison of computational work for the different methods, for the case of parameter set $4$ in Table \ref{table:Reference solution, using MC with $500$ time steps, of Call option price under rBergomi model, for different parameter constellation.}.}
	\label{fig:Complexity plot for MC and MISC for Case set $5$ parameters}
\end{figure}

In this section, we only conduct our numerical experiments for the case the without Richardson extrapolation.  We compare the different methods  in Figure \ref{fig:Complexity plot for MC and MISC for Case set $5$ parameters}, and we determine that both ASGQ and QMC outperform MC, in terms of numerical complexity. In particular,  to achieve a total relative error of approximately  $2\%$, ASGQ  requires	approximately $20\%$ of the work of MC, and  QMC requires approximately $10\%$ of the work of MC. We show more detailed outputs for the methods that are compared in Figure \ref{fig:Complexity plot for MC and MISC for Case set $5$ parameters} in Appendix \ref{appendix:Case of set 4 parameters}.   Similar to the case of set $1$ parameters, illustrated in section \ref{sec:Case of set $2$ parameters_linear}, we believe that the Richardson extrapolation will improve the performance of the ASGQ and QMC methods.   We should also point out that, since we are in the out of the money regime in this case, a fairer comparison of the methods may be done after coupling them with an importance sampling method, so that more points are sampled in the right region of the payoff function.
\FloatBarrier
\FloatBarrier

\section{Conclusions and future work}

In this work,  we propose  novel, fast option pricers,  for options whose underlyings  follow the rBergomi model, as in \cite{bayer2016pricing}.  The new methods  are based on hierarchical deterministic quadrature methods:  i) ASGQ using the same construction as in \cite{haji2016multi}, and ii) the QMC method. Both techniques are coupled with a Brownian bridge construction and the Richardson extrapolation on the weak error.

Given that the only prevalent option, in this context, is to use different variants of the MC method, which is computationally expensive, our main contribution  is to propose a competitive alternative to the MC approach that is based on deterministic quadrature methods.  We believe that we are the first to propose (and design) a pricing method, in the context of rough volatility models, that is based on deterministic quadrature, and that  our approach  opens a new research direction  to investigate the performance of other methods besides MC, for pricing and calibrating under rough volatility models. 

Assuming one targets price estimates with a sufficiently small relative error tolerance, our proposed methods demonstrate substantial computational gains  over the standard MC method, when pricing under the rBergomi model, even for very small values of the Hurst parameter. We show  these gains through our numerical experiments for  different parameter constellations.  We clarify that we do not claim that these gains will hold in the asymptotic regime, i.e.,  for higher accuracy requirements. Furthermore, the use of the Richardson extrapolation is justified in the pre-asymptotic regime, for which our observed experimental results suggest a convergence of order one for the weak error. We emphasize that, to the best of our knowledge, no proper weak error analysis has been done in the rough volatility context, but we claim that both hybrid and exact schemes have  a weak error  of order one, which is justified, at least numerically.  While our focus is on the rBergomi model, our approach is applicable to a wide class of stochastic volatility models, in particular rough volatility models.

In this work, we limit ourselves to compare our novel proposed method against the standard MC. A more systematic comparison against the variant of MC proposed in \cite{mccrickerd2018turbocharging}  can be carried out, but this remains for a future study. Another future research direction is to provide a reliable method for controlling the quadrature error for ASGQ which is,  to the best of our knowledge,  still  an open research problem. This is even more challenging in our context, especially for low values of $H$. We emphasize that the main aim of this work is to illustrate the high potential of the deterministic quadrature, when coupled with hierarchical representations, for pricing options under the rBergomi model. Lastly, we note that accelerating  our novel  approach can be achieved  by using better versions of the ASGQ and QMC methods or an alternative weak approximation of rough volatility models, based on a Donsker type approximation, as recently suggested in \cite{horvath2017functional}. The ASGQ and QMC methods proposed in this work can also be applied in the context of this scheme; in this  case, we only have an $N$-dimensional input rather than a ($2N$)-dimensional input for the integration problem  \eqref{BS_formula_rbergomi_2}, with $N$ being the number of time steps. Further analysis, in particular regarding the Richardson approximation, will be done in future research.

\

\textbf{Acknowledgments}  C. Bayer gratefully acknowledges support from the German Research Foundation (DFG), via the Cluster of Excellence MATH+ (project AA4-2) and the individual grant BA5484/1. This work was supported by the KAUST Office of Sponsored Research (OSR) under Award No. URF/1/2584-01-01 and the Alexander von Humboldt Foundation. C. Ben Hammouda and R. Tempone are members of the KAUST SRI Center for Uncertainty Quantification in Computational Science and Engineering. The authors would like to thank Joakim Beck, Eric Joseph Hall and Erik von Schwerin for their helpful and constructive comments. The authors are also very grateful to the anonymous referees for their valuable comments and suggestions  that greatly contributed to shape the final version of the paper.


\bibliographystyle{plain}
\bibliography{smoothing_rBergomi.bib}

\begin{thebibliography}{10}

\bibitem{abi2019lifting}
Eduardo Abi~Jaber.
\newblock Lifting the {H}eston model.
\newblock {\em Quantitative Finance}, 19(12):1995--2013, 2019.

\bibitem{acworth1998comparison}
Peter~A Acworth, Mark Broadie, and Paul Glasserman.
\newblock A comparison of some {M}onte {C}arlo and quasi {M}onte {C}arlo
  techniques for option pricing.
\newblock In {\em Monte Carlo and Quasi-Monte Carlo Methods 1996}, pages 1--18.
  Springer, 1998.

\bibitem{alos2007short}
Elisa Al{\`o}s, Jorge~A Le{\'o}n, and Josep Vives.
\newblock On the short-time behavior of the implied volatility for
  jump-diffusion models with stochastic volatility.
\newblock {\em Finance and Stochastics}, 11(4):571--589, 2007.

\bibitem{bajgrowicz2015jumps}
Pierre Bajgrowicz, Olivier Scaillet, and Adrien Treccani.
\newblock Jumps in high-frequency data: Spurious detections, dynamics, and
  news.
\newblock {\em Management Science}, 62(8):2198--2217, 2015.

\bibitem{bayer2016pricing}
Christian Bayer, Peter Friz, and Jim Gatheral.
\newblock Pricing under rough volatility.
\newblock {\em Quantitative Finance}, 16(6):887--904, 2016.

\bibitem{bayer2017regularity}
Christian Bayer, Peter~K Friz, Paul Gassiat, Joerg Martin, and Benjamin
  Stemper.
\newblock A regularity structure for rough volatility.
\newblock {\em arXiv preprint arXiv:1710.07481}, 2017.

\bibitem{bayer2018short}
Christian Bayer, Peter~K Friz, Archil Gulisashvili, Blanka Horvath, and
  Benjamin Stemper.
\newblock Short-time near-the-money skew in rough fractional volatility models.
\newblock {\em Quantitative Finance}, pages 1--20, 2018.

\bibitem{bayersmoothing}
Christian Bayer, Markus Siebenmorgen, and R\'aul Tempone.
\newblock Smoothing the payoff for efficient computation of basket option
  pricing.
\newblock {\em Quantitative Finance}, 18(3):491--505, 2018.

\bibitem{bennedsen2016decoupling}
Mikkel Bennedsen, Asger Lunde, and Mikko~S Pakkanen.
\newblock Decoupling the short-and long-term behavior of stochastic volatility.
\newblock {\em arXiv preprint arXiv:1610.00332}, 2016.

\bibitem{bennedsen2017hybrid}
Mikkel Bennedsen, Asger Lunde, and Mikko~S Pakkanen.
\newblock Hybrid scheme for {B}rownian semistationary processes.
\newblock {\em Finance and Stochastics}, 21(4):931--965, 2017.

\bibitem{bergomi2005smile}
Lorenzo Bergomi.
\newblock Smile dynamics {II}.
\newblock {\em Risk}, 18:67–--73, 2005.

\bibitem{biagini2008stochastic}
F.~Biagini, Y.~Hu, B.~{\O}ksendal, and T.~Zhang.
\newblock {\em Stochastic calculus for fractional {B}rownian motion and
  applications}.
\newblock Probability and its Applications. Springer London, 2008.

\bibitem{bungartz2004sparse}
Hans-Joachim Bungartz and Michael Griebel.
\newblock Sparse grids.
\newblock {\em Acta numerica}, 13:147--269, 2004.

\bibitem{caflisch1997valuation}
Russel~E Caflisch, William~J Morokoff, and Art~B Owen.
\newblock {\em Valuation of mortgage backed securities using {B}rownian bridges
  to reduce effective dimension}.
\newblock 1997.

\bibitem{christensen2014fact}
Kim Christensen, Roel~CA Oomen, and Mark Podolskij.
\newblock Fact or friction: Jumps at ultra high frequency.
\newblock {\em Journal of Financial Economics}, 114(3):576--599, 2014.

\bibitem{cools2008belgian}
Ronald Cools and Dirk Nuyens.
\newblock A {B}elgian view on lattice rules.
\newblock In {\em Monte Carlo and Quasi-Monte Carlo Methods 2006}, pages 3--21.
  Springer, 2008.

\bibitem{coutin07introduction}
Laure Coutin.
\newblock An introduction to (stochastic) calculus with respect to fractional
  {B}rownian motion.
\newblock In {\em S{\'e}minaire de Probabilit{\'e}s XL}, pages 3--65. Springer,
  2007.

\bibitem{el2018roughening}
Omar El~Euch, Jim Gatheral, and Mathieu Rosenbaum.
\newblock Roughening {H}eston.
\newblock {\em Available at SSRN 3116887}, 2018.

\bibitem{el2019characteristic}
Omar El~Euch and Mathieu Rosenbaum.
\newblock The characteristic function of rough {H}eston models.
\newblock {\em Mathematical Finance}, 29(1):3--38, 2019.

\bibitem{el2018perfect}
Omar El~Euch, Mathieu Rosenbaum, et~al.
\newblock Perfect hedging in rough {H}eston models.
\newblock {\em The Annals of Applied Probability}, 28(6):3813--3856, 2018.

\bibitem{forde2017asymptotics}
Martin Forde and Hongzhong Zhang.
\newblock Asymptotics for rough stochastic volatility models.
\newblock {\em SIAM Journal on Financial Mathematics}, 8(1):114--145, 2017.

\bibitem{fukasawa2011asymptotic}
Masaaki Fukasawa.
\newblock Asymptotic analysis for stochastic volatility: martingale expansion.
\newblock {\em Finance and Stochastics}, 15(4):635--654, 2011.

\bibitem{gatheral2014volatility_2}
Jim Gatheral, Thibault Jaisson, Andrew Lesniewski, and Mathieu Rosenbaum.
\newblock Volatility is rough, part 2: Pricing.
\newblock Workshop on Stochastic and Quantitative Finance, Imperial College
  London, London, 2014.

\bibitem{gatheral2018volatility}
Jim Gatheral, Thibault Jaisson, and Mathieu Rosenbaum.
\newblock Volatility is rough.
\newblock {\em Quantitative Finance}, 18(6):933--949, 2018.

\bibitem{gatheral2019affine}
Jim Gatheral and Martin Keller-Ressel.
\newblock Affine forward variance models.
\newblock {\em Finance and Stochastics}, pages 1--33.

\bibitem{glasserman2004monte}
Paul Glasserman.
\newblock {\em Monte {C}arlo methods in financial engineering}.
\newblock Springer, New York, 2004.

\bibitem{haji2016multi}
Abdul-Lateef Haji-Ali, Fabio Nobile, Lorenzo Tamellini, and R\'aul Tempone.
\newblock Multi-index stochastic collocation for random {PDE}s.
\newblock {\em Computer Methods in Applied Mechanics and Engineering},
  306:95--122, 2016.

\bibitem{horvath2017functional}
Blanka Horvath, Antoine Jacquier, and Aitor Muguruza.
\newblock Functional central limit theorems for rough volatility.
\newblock {\em Available at SSRN 3078743}, 2017.

\bibitem{imai2004minimizing}
Junichi Imai and Ken~Seng Tan.
\newblock Minimizing effective dimension using linear transformation.
\newblock In {\em Monte Carlo and Quasi-Monte Carlo Methods 2002}, pages
  275--292. Springer, 2004.

\bibitem{jaber2019affine}
Eduardo~Abi Jaber, Martin Larsson, Sergio Pulido, et~al.
\newblock Affine {V}olterra processes.
\newblock {\em The Annals of Applied Probability}, 29(5):3155--3200, 2019.

\bibitem{jacquier2018vix}
Antoine Jacquier, Claude Martini, and Aitor Muguruza.
\newblock On {VIX} futures in the rough {B}ergomi model.
\newblock {\em Quantitative Finance}, 18(1):45--61, 2018.

\bibitem{jacquier2017pathwise}
Antoine Jacquier, Mikko~S Pakkanen, and Henry Stone.
\newblock Pathwise large deviations for the rough {B}ergomi model.
\newblock {\em Journal of Applied Probability}, 55(4):1078--1092, 2018.

\bibitem{mandelbrot1968fractional}
Benoit~B Mandelbrot and John~W Van~Ness.
\newblock Fractional {B}rownian motions, fractional noises and applications.
\newblock {\em SIAM review}, 10(4):422--437, 1968.

\bibitem{marinucci1999alternative}
Domenico Marinucci and Peter~M Robinson.
\newblock Alternative forms of fractional {B}rownian motion.
\newblock {\em Journal of statistical planning and inference},
  80(1-2):111--122, 1999.

\bibitem{mccrickerd2018turbocharging}
Ryan McCrickerd and Mikko~S Pakkanen.
\newblock Turbocharging {M}onte {C}arlo pricing for the rough {B}ergomi model.
\newblock {\em Quantitative Finance}, pages 1--10, 2018.

\bibitem{morokoff1994quasi}
William~J Morokoff and Russel~E Caflisch.
\newblock Quasi-random sequences and their discrepancies.
\newblock {\em SIAM Journal on Scientific Computing}, 15(6):1251--1279, 1994.

\bibitem{moskowitz1996smoothness}
Bradley Moskowitz and Russel~E Caflisch.
\newblock Smoothness and dimension reduction in quasi-{M}onte {C}arlo methods.
\newblock {\em Mathematical and Computer Modelling}, 23(8):37--54, 1996.

\bibitem{neuenkirch2016order}
Andreas Neuenkirch and Taras Shalaiko.
\newblock The order barrier for strong approximation of rough volatility
  models.
\newblock {\em arXiv preprint arXiv:1606.03854}, 2016.

\bibitem{nuyens2014construction}
Dirk Nuyens.
\newblock The construction of good lattice rules and polynomial lattice rules.,
  2014.

\bibitem{picard2011representation}
Jean Picard.
\newblock Representation formulae for the fractional {B}rownian motion.
\newblock In {\em S{\'e}minaire de Probabilit{\'e}s XLIII}, pages 3--70.
  Springer, 2011.

\bibitem{romano1997contingent}
Marc Romano and Nizar Touzi.
\newblock Contingent claims and market completeness in a stochastic volatility
  model.
\newblock {\em Mathematical Finance}, 7(4):399--412, 1997.

\bibitem{sloan1985lattice}
Ian~H Sloan.
\newblock Lattice methods for multiple integration.
\newblock {\em Journal of Computational and Applied Mathematics}, 12:131--143,
  1985.

\bibitem{talay1990expansion}
Denis Talay and Luciano Tubaro.
\newblock Expansion of the global error for numerical schemes solving
  stochastic differential equations.
\newblock {\em Stochastic analysis and applications}, 8(4):483--509, 1990.

\end{thebibliography}

\appendix
\section{Appendix}
\subsection{Case of set $1$ parameters in table \ref{table:Reference solution, using MC with $500$ time steps, of Call option price under rBergomi model, for different parameter constellation.}}\label{appendix:Case of set 1 parameters}

\begin{table}[h!]
	\begin{small}
		\centering
		\begin{tabular}{l*{6}{c}r}
			\toprule[1.5pt]
			Method & & Steps  & &     \\
			\hline
			& $1-2$ & $2-4$ & $4-8$   & $8-16$  \\
			\hline	
			QMC + level $1$ of  Richardson extrapolation &$\underset{(0.96,0.91)}{\mathbf{1.87}}$  & $\underset{(0.07,0.09)}{\mathbf{0.16}}$ & $\underset{(0.015,0.018)}{\mathbf{0.033}}$ & $\underset{(0.002,0.002)}{\mathbf{0.0044}}$  \\
			M(\# QMC samples) & $128$  & $8192$ & $
			131072$ & $
			2097152$ \\
			\hline
			
			MC + level $1$ of  Richardson extrapolation &$\underset{(0.96,0.92)}{\mathbf{1.88}}$  & $\underset{(0.07,0.07)}{\mathbf{0.14}}$ & $\underset{(0.015,0.015)}{\mathbf{0.03}}$  & $\underset{(0.002,0.0024)}{\mathbf{0.0044}}$\\
			M(\# MC samples) & $4 \times 10$  & $8 \times 10^3$ & $16 \times 10^4$ & $5 \times 10^5$  \\
			\bottomrule[1.25pt]
		\end{tabular}
		\caption{Total relative error of MC and randomized QMC coupled with the Richardson extrapolation (level $1$), to compute the call option price  for different numbers of time steps. The values between parentheses correspond to the different errors contributing to the total relative error: the bias and the statistical error estimates. The number of MC and QMC samples, $M$, are chosen to satisfy \eqref{optimal_number_samples}.}
		\label{Total  error of MISC and MC to compute Call option price of the different tolerances for different number of time steps. Case set $2$ parameters, with Richardson extrapolation(level $1$). The numbers between parentheses are the corresponding absolute errors,relative}
	\end{small}
\end{table}
\FloatBarrier

\begin{table}[h!]
	\centering
	\begin{tabular}{l*{6}{c}r}
		\toprule[1.5pt]
		Method & &   & Steps & &     \\
		\hline
		& $1-2$ & $2-4$ & $4-8$ & $8-16$   \\
		\hline	
		QMC + level $1$ of  Richardson extrapolation  &$0.018$ & $2$  & $18$  & $333$   \\
		\hline	
		MC + level $1$ of  Richardson extrapolation &$
		0.0012$ & $12$  & $152$  & $4400$ \\
		\bottomrule[1.25pt]
	\end{tabular}
	\caption{Comparison of the computational time (in seconds) of  MC and randomized QMC coupled with  the Richardson extrapolation (level $1$) to compute the call option price of the rBergomi model for different numbers of time steps. The average MC CPU time is computed over $100$ runs.}
	\label{Comparsion of the computational time of  MC and MISC, using Richardson extrapolation (level $1$), used to compute Call option price of rBergomi model for different number of time steps. Case set $2$ parameters,linear}
\end{table}

\FloatBarrier

\begin{table}[!h]
	\begin{small}
		\centering
		\begin{tabular}{l*{6}{c}r}
			\toprule[1.5pt]
			Method & & Steps  & &     \\
			\hline
			& $1-2-4$ & $2-4-8$  \\
			\hline
			
			ASGQ + level $2$ of  Richardson extrapolation ($\text{TOL}_{\text{ASGQ}}=10^{-1}$)  & $\underset{(0.24,0.30)}{\mathbf{ 0.54
			}}$ & $\underset{(0.006,0.107)}{\mathbf{ 0.113}}$ \\
			ASGQ + level $2$ of  Richardson extrapolation ($\text{TOL}_{\text{ASGQ}}=5.10^{-2}$)  & $\underset{(0.24,0.25)}{\mathbf{   0.49
			}}$ & $\underset{(0.006,0.003)}{\mathbf{ 0.009} }$  \\
			\bottomrule[1.25pt]
		\end{tabular}
		\caption{Total relative error of ASGQ, coupled with  the Richardson extrapolation (level $2$), to compute the call option price for different numbers of time steps.  The values between parentheses correspond to the different errors contributing to the total relative error: the bias and quadrature errors.}
		\label{Total  error of MISC and MC to compute Call option price of the different tolerances for different number of time steps. Case set $2$ parameters, with Richardson extrapolation(level $2$). The numbers between parentheses are the corresponding absolute errors,linear}
	\end{small}
\end{table}
\FloatBarrier

\begin{table}[!h]
	\centering
	\begin{tabular}{l*{6}{c}r}
		\toprule[1.5pt]
		Method & & Steps  & &     \\
		\hline
		& $1-2-4$ & $2-4-8$   \\
		\hline
		
		ASGQ + level $2$ of  Richardson extrapolation ($\text{TOL}_{\text{ASGQ}}=10^{-1}$)  & $0.2$ & $2$ &   \\
		ASGQ + level $2$ of  Richardson extrapolation ($\text{TOL}_{\text{ASGQ}}=5.10^{-2}$)  & $0.5$ & $74$  \\
		\bottomrule[1.25pt]
	\end{tabular}
	\caption{Comparison of the computational time (in seconds) of ASGQ coupled with the Richardson extrapolation (level $2$) to compute the call option price of the rBergomi model for different numbers of time steps.}
	\label{Comparsion of the computational time of  MC and MISC, using Richardson extrapolation (level $2$), used to compute Call option price of rBergomi model for different number of time steps. Case set $2$ parameters,linear}
\end{table}
\FloatBarrier

\subsection{Case of set $2$ parameters in table \ref{table:Reference solution, using MC with $500$ time steps, of Call option price under rBergomi model, for different parameter constellation.} }
\label{appendix:Case of set $2$ parameters}

\FloatBarrier

\begin{table}[h!]
	\begin{small}
		\centering
		\begin{tabular}{l*{6}{c}r}
			\toprule[1.5pt]
			Method & & Steps  & &     \\
			\hline		
			& $2$ & $4$ & $8$ & $16$  \\
			\hline
			ASGQ ($\text{TOL}_{\text{ASGQ}}=10^{-1}$)  &  $\underset{(0.02,0.01)}{\mathbf{0.03}}$ & $\underset{(0.008,0.014)}{\mathbf{0.022}}$& $\underset{(0.004,0.018)}{\mathbf{ 0.022}}$ & $\underset{(0.001,0.016)}{\mathbf{ 0.017}}$   \\
			
			ASGQ ($\text{TOL}_{\text{ASGQ}}=10^{-2}$)  &  $\underset{(0.02,0.01)}{\mathbf{0.03}}$ & $\underset{(0.008,0.009)}{\mathbf{0.017}}$& $\underset{(0.004,0.004)}{\mathbf{ 0.008}}$ & $\underset{(0.001,4e-04)}{\mathbf{ 0.001}}$  \\
			\hline
			QMC   & $\underset{(0.02,0.02)}{\mathbf{0.04}}$  &  $\underset{(0.008,0.009)}{\mathbf{0.017}}$  & $\underset{(0.004,0.004)}{\mathbf{0.008}}$ & $\underset{(0.001,0.001)}{\mathbf{0.002}}$  \\	
			M(\# QMC samples) 	& $4096$  &  $8192$  & $32768$ & $262144$ \\
			\hline
			MC    & $\underset{(0.02,0.02)}{\mathbf{0.04}}$  &  $\underset{(0.008,0.008)}{\mathbf{0.016}}$  & $\underset{(0.004,0.003)}{\mathbf{0.007}}$ & $\underset{(0.001,0.001)}{\mathbf{0.002}}$  \\	
			M(\# MC samples) 	& $16 \times 10^3$  &  $8 \times 10^4$  & $ 4 \times 10^5$ & $4 \times 10^6$  \\
			\bottomrule[1.25pt]
		\end{tabular}
		\caption{Total relative error of the different methods without the Richardson extrapolation, to compute the call option price for different numbers of time steps. The values between parentheses correspond to the different errors contributing to the total relative error; for ASGQ, we report the bias and quadrature errors, and for MC and QMC, we report the bias and the statistical error estimates. The number of MC and QMC  samples, $M$, are chosen to satisfy \eqref{optimal_number_samples}.}
		\label{Total error of MISC and MC to compute Call option price of the different tolerances for different number of time steps. Case set 3, without Richardson extrapolation. The numbers between parentheses are the corresponding absolute errors.}
	\end{small}
\end{table}

\FloatBarrier
\begin{table}[h!]
	\centering
	\begin{tabular}{l*{6}{c}r}
		\toprule[1.5pt]
		Method & & Steps  & &     \\
		\hline	
		& $2$ & $4$ & $8$ & $16$ &   \\
		\hline
		ASGQ ($\text{TOL}_{\text{ASGQ}}=10^{-1}$)  & $0.1$ & $0.1$ & $0.2$ & $0.8$ \\
		ASGQ ($\text{TOL}_{\text{ASGQ}}=10^{-2}$)  & $0.1$ & $0.5$ & $8$ & $92$ \\
		\hline
		QMC method   & $ 0.3$  & $ 0.7$  & $ 3.25$ & $ 27$  \\	
		\hline
		MC method   & $ 0.6$  & $  6.4$  & $  66$ & $ 1976$  \\	
		\bottomrule[1.25pt]
	\end{tabular}
	\caption{Comparison of the computational time (in seconds) of  the different methods to compute the call option price of the rBergomi model for different numbers of time steps. The average  MC CPU time is computed over $100$ runs. }
	\label{Comparsion of the computational time of  MC and MISC, used to compute Call option price of rBergomi model for different number of time steps. Case set3}
\end{table}

\FloatBarrier

\subsection{Case of set $3$ parameters in table \ref{table:Reference solution, using MC with $500$ time steps, of Call option price under rBergomi model, for different parameter constellation.}}\label{appendix:Case of set 3 parameters}

\FloatBarrier

\begin{table}[h!]
	\begin{small}
		\centering
		\begin{tabular}{l*{6}{c}r}
			\toprule[1.5pt]
			Method & & Steps  & &     \\
			\hline	
			& $2$ & $4$ & $8$ & $16$  \\
			\hline
			
			ASGQ ($\text{TOL}_{\text{ASGQ}}=10^{-1}$)  &  $\underset{(0.006,0.002)}{\mathbf{0.008}}$ & $\underset{(0.004,0.005)}{\mathbf{0.009}}$& $\underset{(0.003,0.005)}{\mathbf{ 0.008}}$ & $\underset{(0.002,0.007)}{\mathbf{ 0.009}}$   \\
			
			ASGQ ($\text{TOL}_{\text{ASGQ}}=10^{-2}$)  &  $\underset{(0.006,0.002)}{\mathbf{0.008}}$ & $\underset{(0.004,0.005)}{\mathbf{0.009}}$& $\underset{(0.003,0.002)}{\mathbf{ 0.005}}$ & $\underset{(0.002,1e-04)}{\mathbf{ 0.002}}$  \\
			ASGQ ($\text{TOL}_{\text{ASGQ}}=10^{-3}$)  &  $\underset{(0.006,0.002)}{\mathbf{0.008}}$& $\underset{(0.004,0.002)}{\mathbf{0.006}}$& $\underset{(0.003,1e-04)}{\mathbf{0.003}}$  & $\underset{(0.002,1e-04)}{\mathbf{ 0.002}}$  \\
			ASGQ ($\text{TOL}_{\text{ASGQ}}=10^{-4}$)  &  $\underset{(0.006,4e-04)}{\mathbf{0.006}}$ & $\underset{(0.004,2e-04)}{\mathbf{0.004}}$& $\underset{(0.003,1e-04)}{\mathbf{0.003}}$ & $\mathbf{ -}$ \\

			\hline
			QMC    & $\underset{(0.006,0.009)}{\mathbf{0.015}}$  & $\underset{(0.004,0.004)}{ \mathbf{0.008}}$  & $\underset{(0.003,0.0036)}{\mathbf{0.0066}}$ & $\underset{(0.002,0.002)}{\mathbf{0.004}}$  \\	
			M(\# QMC samples) 	& $2^3 \times 2^{10}= 8192$  &  $2^3 \times 2^{11}=  16384$ &  $2^3 \times 2^{12}= 32768$ & $2^3 \times 2^{13}= 65536	$  \\
			\hline
			MC    & $\underset{(0.006,0.005)}{\mathbf{0.01}}$  & $\underset{(0.004,0.004)}{ \mathbf{0.008}}$  & $\underset{(0.003,0.003)}{\mathbf{0.006}}$ & $\underset{(0.002,0.002)}{\mathbf{0.004}}$  \\	
			M(\# MC samples) 	& $8 \times 10^4$  & $16 \times 10^4$  & $24 \times 10^4$ & $32 \times 10^4$  \\
			\bottomrule[1.25pt]
		\end{tabular}
		\caption{Total relative error of  the different methods without  the Richardson extrapolation,  to compute the call option price  for different numbers of time steps. The values between parentheses correspond to the different errors contributing to the total relative error; for ASGQ, we report the bias and quadrature errors, and for MC and QMC, we report the bias and the statistical error estimates. The number of MC and QMC  samples, $M$, are chosen to satisfy \eqref{optimal_number_samples}.}
		\label{Total error of MISC and MC to compute Call option price of the different tolerances for different number of time steps. Case set 4, without Richardson extrapolation. The numbers between parentheses are the corresponding absolute errors.}
	\end{small}
\end{table}

\FloatBarrier
\begin{table}[h!]
	\centering
	\begin{tabular}{l*{6}{c}r}
		\toprule[1.5pt]
		Method & & Steps  & &     \\
		\hline	
		& $2$ & $4$ & $8$ & $16$ &   \\
		\hline
		ASGQ ($\text{TOL}_{\text{ASGQ}}=10^{-1}$)  & $0.1$ & $0.1$ & $0.1$ & $1$ \\
		ASGQ ($\text{TOL}_{\text{ASGQ}}=10^{-2}$)  & $0.1$ & $0.15$ & $9$ & $112$ \\
		ASGQ ($\text{TOL}_{\text{ASGQ}}=10^{-3}$)  & $0.2$ & $2$ & $27$ & $2226$ \\
		ASGQ ($\text{TOL}_{\text{ASGQ}}=10^{-4}$)  & $1$ & $6$ & $136$ & $-$\\
		\hline
		QMC method    & $0.65$  & $ 1.4$  & $  3.25$ & $ 7.5
		$  \\		
		\hline
		MC method   & $4$  & $ 12$  & $  40$ & $ 160
		$  \\	
		\bottomrule[1.25pt]
	\end{tabular}
	\caption{Comparison of the computational time (in seconds) of the different methods to compute the call option price of the rBergomi model for different numbers of time steps. The average  MC CPU time is computed over $100$ runs. }
	\label{Comparsion of the computational time of  MC and MISC, used to compute Call option price of rBergomi model for different number of time steps. Case set4}
\end{table}

\FloatBarrier

\subsection{Case of set $4$ parameters in table \ref{table:Reference solution, using MC with $500$ time steps, of Call option price under rBergomi model, for different parameter constellation.}}\label{appendix:Case of set 4 parameters}
\FloatBarrier
\begin{table}[h!]
	\begin{small}
		\centering
		\begin{tabular}{l*{6}{c}r}
			\toprule[1.5pt]
			Method & & Steps  & &     \\
			\hline	
			& $2$ & $4$ & $8$ & $16$  \\
			\hline
			
			ASGQ ($\text{TOL}_{\text{ASGQ}}=10^{-1}$)  & $\underset{(0.07,0.05)}{\mathbf{0.09}}$ & $\underset{(0.03,0.04)}{\mathbf{0.07}}$& $\underset{(0.02,0.05)}{\mathbf{ 0.07}}$ & $\underset{(0.01,2e-04)}{\mathbf{ 0.06}}$   \\
			
			ASGQ ($\text{TOL}_{\text{ASGQ}}=10^{-2}$)  &  $\underset{(0.07,5e-04)}{\mathbf{0.09}}$& $\underset{(0.03,0.04)}{\mathbf{0.07}}$& $\underset{(0.02,3e-04)}{\mathbf{ 0.02}}$ & $\underset{(0.01,2e-04)}{\mathbf{ 0.02}}$  \\
			ASGQ ($\text{TOL}_{\text{ASGQ}}=10^{-3}$)  &  $\underset{(0.07,5e-04)}{\mathbf{0.07}}$& $\underset{(0.03,4e-04)}{\mathbf{0.03}}$& $\underset{(0.02,3e-04)}{\mathbf{0.02}}$  & $\underset{(0.01,2e-04)}{\mathbf{ 0.01}}$  \\
			\hline
			QMC     & $\underset{(0.07,0.085)}{\mathbf{0.155}}$  & $\underset{(0.03,0.04)}{\mathbf{0.07}}$  & $\underset{(0.02,0.019)}{\mathbf{0.039}}$ & $\underset{(0.01,0.01)}{\mathbf{0.02}}$  \\		
			M(\# QMC samples)   & $2^3 \times 2^8=2048$  &  $2^3 \times 2^9=4096$  &  $2^3 \times 2^{11}=16384$ &  $2^3 \times 2^{12}=32768$ \\
			\hline
			MC    & $\underset{(0.07,0.07)}{\mathbf{0.14}}$  & $\underset{(0.03,0.04)}{\mathbf{0.07}}$  & $\underset{(0.02,0.02)}{\mathbf{0.04}}$ & $\underset{(0.01,0.01)}{\mathbf{0.02}}$  \\		
			M(\# MC samples)   & $24 \times 10^2$  & $8 \times 10^3$  & $32 \times 10^3$ & $8 \times 10^4$  \\		
			\bottomrule[1.25pt]
		\end{tabular}
		\caption{Total relative error of the different methods without the Richardson extrapolation, to compute the call option price  for different numbers of time steps. The values between parentheses correspond to the different errors contributing to the total relative error; for ASGQ, we report the bias and quadrature errors, and for MC and QMC, we report the bias and the statistical error estimates. The number of MC and QMC samples, $M$, are chosen to satisfy \eqref{optimal_number_samples}.}
		\label{Total error of MISC and MC to compute Call option price of the different tolerances for different number of time steps. Case set 5, without Richardson extrapolation. The numbers between parentheses are the corresponding absolute errors.}
	\end{small}
\end{table}

\FloatBarrier
\begin{table}[h!]
	\centering
	\begin{tabular}{l*{6}{c}r}
		\toprule[1.5pt]
		Method & & Steps  & &     \\
		\hline	
		& $2$ & $4$ & $8$ & $16$ &   \\
		\hline
		ASGQ ($\text{TOL}_{\text{ASGQ}}=10^{-1}$)  & $0.1$ & $0.1$ & $0.2$ & $0.5$ \\
		ASGQ ($\text{TOL}_{\text{ASGQ}}=10^{-2}$)  & $0.1$ & $0.1$ & $8$ & $97$ \\
		ASGQ ($\text{TOL}_{\text{ASGQ}}=10^{-3}$)  & $0.7$ & $4$ & $26$ & $1984$ \\
		\hline
		QMC method    & $ 0.17 $  & $  0.35$  & $ 1.6$ & $ 4$  \\	
		MC method   & $ 0.08 $  & $  0.6$  & $ 5.6$ & $ 40$  \\	
		\bottomrule[1.25pt]
	\end{tabular}
	\caption{Comparison of the computational time (in seconds) of the different methods to compute the call option price of rBergomi model for different numbers of time steps. The average  MC CPU time is computed over $100$ runs. }
	\label{Comparsion of the computational time of  MC and MISC, used to compute Call option price of rBergomi model for different number of time steps. Case set5}
\end{table}

\FloatBarrier

\end{document}